%% file: apssamp.tex
\documentclass[%
 reprint,
 amsmath,amssymb,
 aps,
prb,
]{revtex4-2}
\usepackage{amsmath}
\usepackage{mathdots}
\usepackage{amssymb}
\usepackage[LGR,T1]{fontenc}
\usepackage[utf8]{inputenc}
\usepackage{textalpha}
\usepackage{graphicx}% Include figure files
\usepackage{dcolumn}% Align table columns on decimal point
\usepackage{bm}% bold math
\usepackage[thinlines]{easytable}
\usepackage[unicode=true,pdfusetitle,
 bookmarks=true,bookmarksnumbered=false,bookmarksopen=false,
 breaklinks=true,pdfborder={0 0 0},pdfborderstyle={},backref=false,colorlinks=true]{hyperref}
 \usepackage{mathtools}
\usepackage[table]{xcolor}
\usepackage{amssymb}
\usepackage{pifont}% 
\usepackage{bbding}
\usepackage{pgfplots}
\pgfplotsset{compat=1.18} % or your preferred version
\usepackage{tikz}
\usetikzlibrary{fpu,datavisualization,intersections,arrows,decorations.pathmorphing,backgrounds,positioning,fit,petri}

\newlength\fheight
\newlength\fwidth

\newcommand{\xmark}{\ding{55}}% for the "x" mark
\newcommand\SmallMatrix[1]{{%
\footnotesize\arraycolsep=0.3\arraycolsep\ensuremath{\begin{pmatrix}#1\end{pmatrix}}}}

\begin{document}

\preprint{APS/123-QED}

\title{Defects Potentials for Two-Dimensional Topological Materials }% Force line breaks with \\

\author{Yuval Abulafia}
\affiliation{
 Technion - Israel Institute of Technology
}%Lines break automatically or can be forced with \\
\author{Amit Goft}%
\affiliation{
 Technion - Israel Institute of Technology
}
\author{Nadav Orion}
\affiliation{
 Technion - Israel Institute of Technology
}
\author{Eric Akkermans}
 \email{eric@physics.technion.ac.il}
%affiliation{%
 %Authors' institution and/or address
%}%


\affiliation{
 Technion - Israel Institute of Technology
}%


\date{\today}% It is always \today, today,

\begin{abstract}
For non-topological quantum materials, introducing defects can significantly alter their properties by modifying symmetry and generating a nonzero analytical index, thus transforming the material into a topological one. We present a method to construct the potential matrix configuration with the purpose of obtaining a non-zero analytical index, akin to a topological invariant like a winding or Chern number. We establish systematic connections between these potentials, expressed in the continuum limit, and their initial tight-binding model description.
We apply our method to graphene with an adatom, a vacancy, and both as key examples illustrating our comprehensive description. This method enables analytical differentiation between topological and non-topological zero-energy modes and allows for the construction of defects that induce topology.
\end{abstract}

\maketitle

%\tableofcontents

\section{\label{sec:level1}Introduction
\protect}

The search for materials having topological features has become a multifold and active topic of unabated interest \cite{Wieder2022,Hasan2010}. An important goal is to find or synthesize new families of quantum materials with properly tailored specific functionalities. Among possible examples, we may cite enhanced electric transport reducing electrical resistivity \cite{Khan2025}, the building of distinguishable qubits based on topological excitations \cite{Kitaev2003,Fu2008}, and the enhancement of the figure of merit in the search for thermoelectric materials \cite{Takahashi2010}. A common denominator in all these cases is the search for specific quantum states enjoying high immunity to perturbations, e.g. disorder and surface scattering \cite{Kane&Mele2025,Xu&Moore2006}, decoherence, and coupling to other degrees of freedom, as can be granted from topological protection. The purpose of this paper is to show that it is possible to turn 
families of two-dimensional  
non-topological materials into topological by means of properly engineered potentials. 

In materials with weak electronic interactions, topological features are deduced from two anti-unitary and one chiral symmetry. These features are measured using topological quantities like Chern or winding numbers \cite{TKNN1982,KOHMOTO1985}. Time-reversal and particle-hole symmetries typically identify these anti-unitary symmetries, resulting in ten families of crystalline materials known as the tenfold classification \cite{Altland1997, Schnyder2008, Kitaev2009}. These families may exhibit desired topological characteristics or not, based on spatial dimensions $d$. 
A noteworthy extension has been proposed  \cite{Teo2010,Chiu2016,Goft2023} that broadens this tenfold classification beyond crystals, making it possible to compute topological features without a Brillouin zone. 
There, it has been shown that defects or textures, while destroying translation symmetry, still preserve the general form of the tenfold classification by simply replacing the spatial dimension $d$ by a co-dimension $\delta = d -D$ \cite{Teo2010}, where the new dimension $D$ accounts for the specific nature of the defect, i.e., a point, a line, or a texture in $d$-dimensions.

While the study of defects in quantum materials is well-documented \cite{Kelly1998, Pereira2006, Amara2007, Pereira2008, Palacios2008, Peres2009A, Peres2009, Dutreix2013, Dutreix2016, Nanda2013, Ovdat2020,Ugeda2010, Ovdat2017}, there persists some confusion about their topological features. We aim to systematically identify and exploit defect potential aspects that convert non-topological materials into topological types. We describe topology using the analytical index, an alternative to Chern or winding numbers in the tenfold classification. This index, equivalent to the topological invariant, serves as a complementary tool for analysis. 
This paper offers practical examples using 2D $(d=2)$ materials, highlighting vacancies and adatoms.  

Pseudo-spin degrees of freedom are crucial for noting topological features in weakly interacting fermions. In two-dimensional materials, these pseudo-spin degrees of freedom are offered by the lattice's bipartite structure and fermion doubling, resulting in pairs of independent valleys or Dirac points. Their role in the existence of topological features is presented in section \ref{sec:level2}. Then, starting from tight-binding models, we obtain effective first quantized Hamiltonians defined in a Hilbert space that combines orbital and the new pseudo-spin degrees of freedom. This effective approach is systematically used in section \ref{sec:potentials}.
This methodology is implemented using a honeycomb lattice model, effectively describing graphene. Section \ref{sec:TopoZM} examines topological zero modes using the analytical index, providing an alternative to Berry curvature-based topological number calculations requiring translation symmetry and a Brillouin zone. Our approach enables straightforward identification and calculation of topological zero energy states. In Section \ref{sec:level4}, we utilize this method to efficiently compute Green's functions, enabling the evaluation of thermodynamic and transport properties like the electronic density and electrical conductance, as well as properties like Friedel oscillations \cite{Mariani2007, Cheianov2006}.  

\section{\label{sec:level2} Bipartite Lattices - Fermion Doubling and Degrees of Freedom  
}

We consider tight-binding models for free fermions in periodic and bipartite lattices and discuss the existence of anti-unitary symmetries and topology. Bipartite lattices intriguingly relate to topology, influencing physical properties like electrical transport amid bipartite-preserving disorder  \cite{gade1991n,gade1993anderson, FABRIZIO2000542,wu2007weak,ostrovsky2006electron,tikhonenko2008weak}.   

Bipartite lattices with $2N$ sites are divided into two sublattices, $A$ and $B$. As shown in fig \ref{tbl:lattice_chiral_bipatite_valley}, examples include honeycomb (e.g., graphene) \cite{Wallace1947}, brickwall \cite{Pereira2009strain_graphene,Montambaux2009}, and Lieb lattices \cite{Lieb1989}. Although Lieb lattices split into three sublattices, they remain bipartite.

Tight-binding Hamiltonians for free fermions on bipartite lattices consist of hopping terms and on-site potentials. We assign pairs of operators $(a_{\boldsymbol{R}_j}^{\dagger}$, $b_{\boldsymbol{R}_j}^{\dagger})$ at each lattice site ${\boldsymbol{R}_j}$ to identify intra-sublattice terms, $a^{\dagger}_{\boldsymbol{R}_i}a_{\boldsymbol{R}_j}^{\vphantom{\ast}}$ or $b^{\dagger}_{\boldsymbol{R}_i}b_{\boldsymbol{R}_j}^{\vphantom{\ast}}$, for on-site energy when $\boldsymbol{R}_i= \boldsymbol{R}_j$, and inter-sublattice terms, $a^{\dagger}_{{\boldsymbol{R}}_i}b_{{\boldsymbol{R}}_j}^{\vphantom{\ast}}$, for cross-lattice hopping. The generic $2N \times 2N$ Hamiltonian can be written in a sublattice basis,
\begin{equation}\label{eq:ab_hamiltonian}
   \mathcal{H}=\left(\begin{array}{cc}
a^{\dagger} & b^{\dagger}\end{array}\right)\left(\begin{array}{cc}
H_{aa} & H_{ba}^\dagger\\
H_{ba} & H_{bb}
\end{array}\right)\left(\begin{array}{c}
a\\
b
\end{array}\right)
\end{equation}
with $a\equiv \left(a_{\boldsymbol{R}_1},...,a_{\boldsymbol{R}_j},..,a_{\boldsymbol{R}_N}\right)^T$ and $b\equiv \left(b_{\boldsymbol{R}_1},...,b_{\boldsymbol{R}_j},..,b_{\boldsymbol{R}_N}\right)^T$ representing all $N$ sites in sublattices 
$A$ and $B$, respectively. Defining the matrix, 
\begin{equation}\label{eq:ab_hamiltonian1}
  H \equiv  \left(\begin{array}{cc}
H_{aa} & H_{ba}^\dagger\\
H_{ba} & H_{bb}
\end{array}\right) \, ,
\end{equation}
helps to identify $N \times N$ submatrices 
$H_{aa}$ and $H_{bb}$ describing intra-sublattice terms, while off-diagonal $N \times N$ submatrices  $H_{ab}$ and $H_{ba}= H_{ab}^\dagger$ represent inter-sublattice terms. Recognizing this separation helps in understanding the symmetries underlying topological features. For example, particle-hole is preserved if all on-site potentials are constant and the hopping energy is taken to be non-zero only between nearest neighbor sites, hence connecting the two different sublattices $A$ and  $B$.

By providing a partition between intra- and inter-sublattice terms, the bipartite nature of the lattice acts as a binary degree of freedom. For non-Bravais lattices, the two types of terms it involves are directly related to the chiral symmetry \cite{Coulson&Rushbrooke1940,RIGBY1979,Xiao2024}  defined by the anti-commutation relation, 
\begin{equation}\label{eq:chiral_symmetry}
\left\{\sigma_3,{H}\right\}=0,
\end{equation} where $\sigma_3$ is the $2N \times 2N$ generalization of the $(2 \times 2)$ $\sigma_z$ Pauli matrix. When ${H}_{aa}={H}_{bb}=0$, the Hamiltonian is chiral symmetric, 
allowing the identification of zero-energy states, or zero modes. Certain zero modes result from topological constraints, which are further examined in section \ref{sec:TopoZM}.

%%%%%%%%%%%%%%%  Figure %%%%%%%%%%%%%%%%%
%%%%%%%%%%%%%%%%% %%%%%%%%%%%%%%
\begin{figure}[ht]
    \centering
    \begin{tabular}{cc}
\includegraphics[width=0.20\textwidth,bb=170bp 175bp 480bp 400bp,clip]{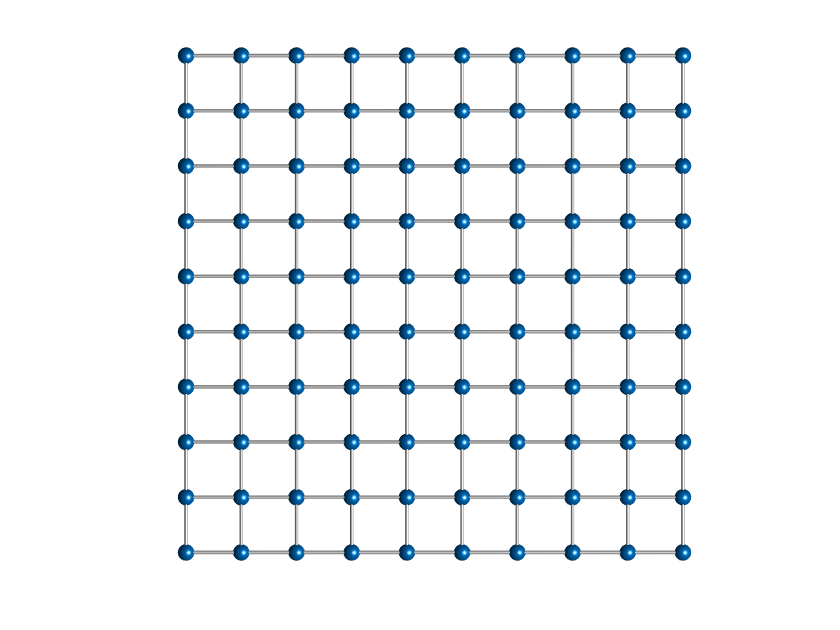}
      &
        \:\: 
\includegraphics[width=0.20\textwidth,bb=195bp 190bp 490bp 400bp,clip]{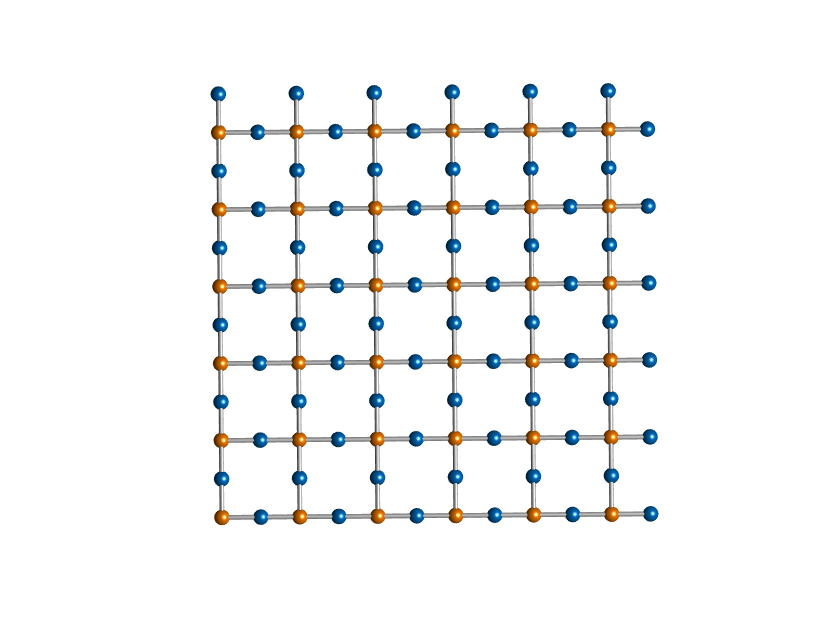} 
      \\
      \small (a) Square & \small (b) Lieb
      \vspace{5pt}
      \\
\includegraphics[width=0.20\textwidth,bb=190bp 185bp 470bp 387bp,clip]{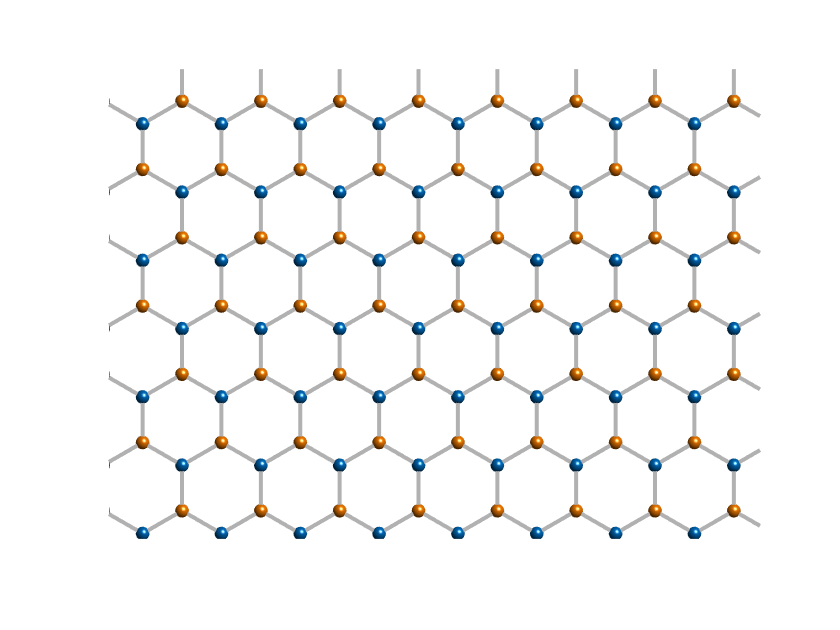}
      & 
     \:\: \includegraphics[width=0.20\textwidth,bb=130bp 120bp 410bp 322bp,clip]{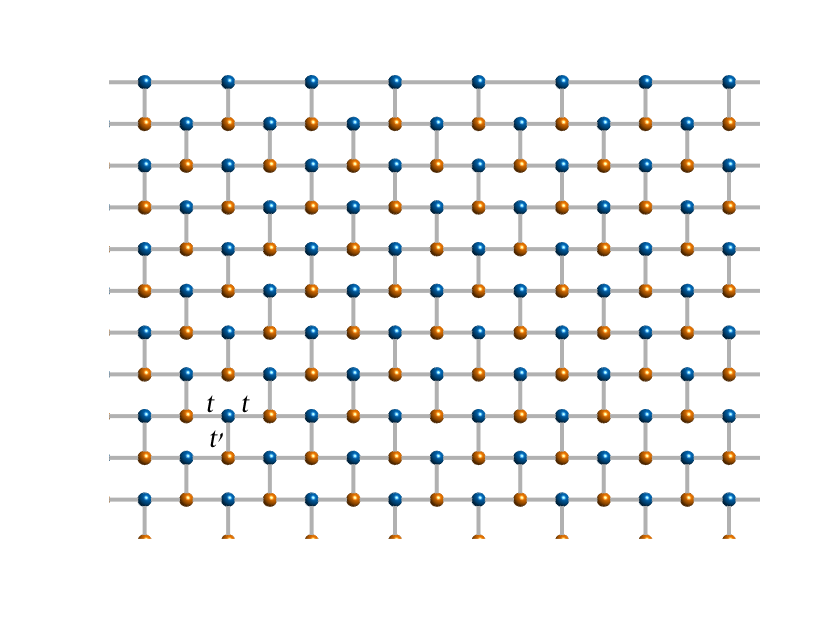}
     \\
    \small (c) Honeycomb & \small (d) Brickwall
      \\

	\end{tabular}
	\caption{Bipartite lattice examples include: The square lattice, a Bravais lattice with a single atom per unit cell; honeycomb and brickwall lattices, each with two atoms per unit cell; and the Lieb lattice, formed by augmenting a square lattice with a central site on every bond, resulting in a unit cell with three sites—one original and two neighboring.
 }
	\label{figure:symmetries in lattices}
\end{figure}
 
\begin{table}[ht]
  \centering
  \begin{tabular}{ | >{\centering\arraybackslash}m{1.65cm} 
                  | >{\centering\arraybackslash}m{1.3cm} 
                  | >{\centering\arraybackslash}m{1.4cm} 
                  | >{\centering\arraybackslash}m{1.5cm} 
                  | >{\centering\arraybackslash}m{1.7cm} | }
    \hline
      Lattice \vspace{1mm} &  \vspace{1mm}  Bipartite  \vspace{-2.7mm} &   Fermion doubling  
     &  Chiral symmetry & 
  Band structure \\ \hline
    \centering Square 
    &
    \centering \checkmark 
    &  \centering \xmark
    & \centering\xmark
    &
     \begin{minipage}{.1\textwidth}
\centering\includegraphics[width=8mm, height=11mm]{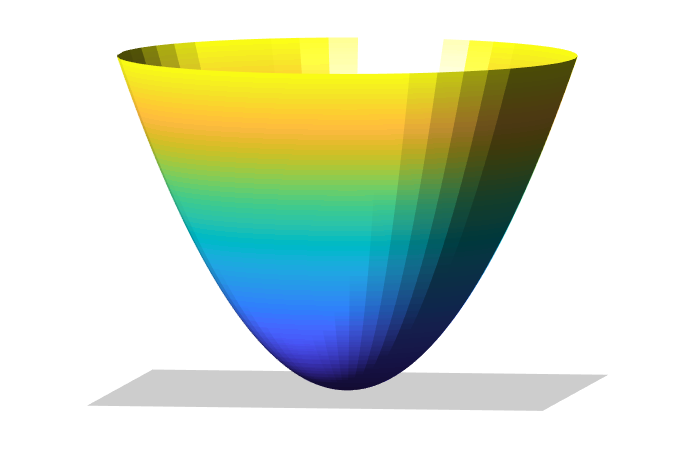}
    \end{minipage}
    \\ \hline
    \centering Lieb 
    & \centering  \checkmark
    &\centering\xmark
    &\centering\checkmark
    &
   \begin{minipage}{.1\textwidth}
\centering\includegraphics[width=8mm, height=10mm]{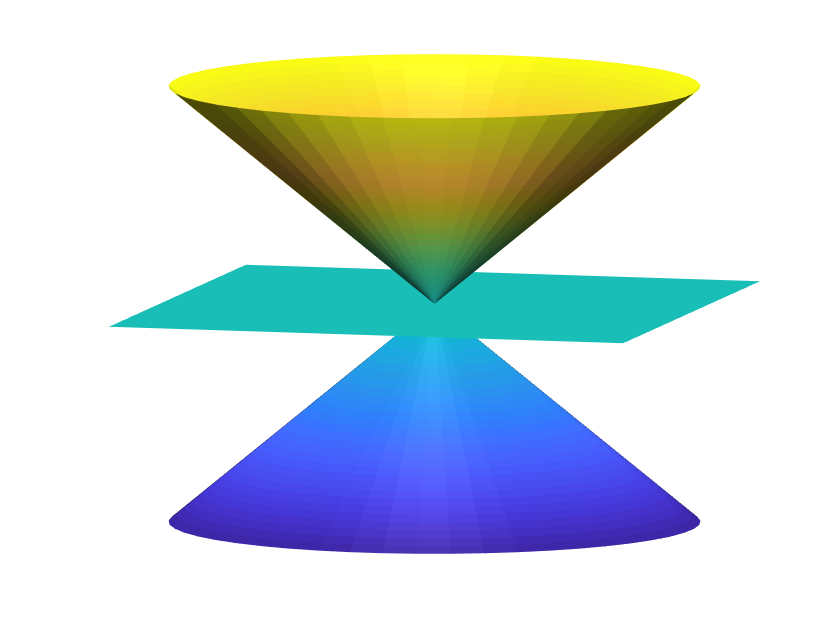}
    \end{minipage}
    \\ \hline
    \centering Honeycomb
    & \centering \checkmark
    & \centering\checkmark
    & \centering\checkmark
    & \vspace{0.1mm}\begin{minipage}{.1\textwidth}
\centering\includegraphics[width=12mm, height=10mm]{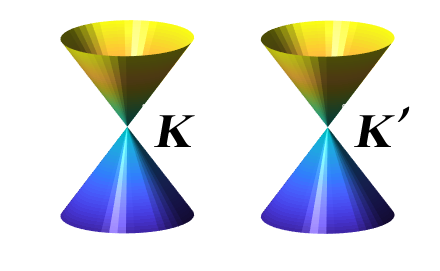}
    \end{minipage}
    \\ \hline
    \vspace{1mm}
  \centering Brickwall  $t'/t<2$  & 
 \centering \checkmark
    & 
\centering\checkmark
& \centering\checkmark
    &\begin{minipage}{.1\textwidth}
\centering\includegraphics[width=12mm, height=10mm]{pics/graphene_energy.png}\end{minipage}
\\ \hline
\centering 
\vspace{1mm}
Brickwall  $t'/t>2$ &\centering \checkmark &
\centering\text{\xmark}
& \centering \checkmark &\begin{minipage}{.1\textwidth}
\centering\includegraphics[width=8mm, height=10mm]{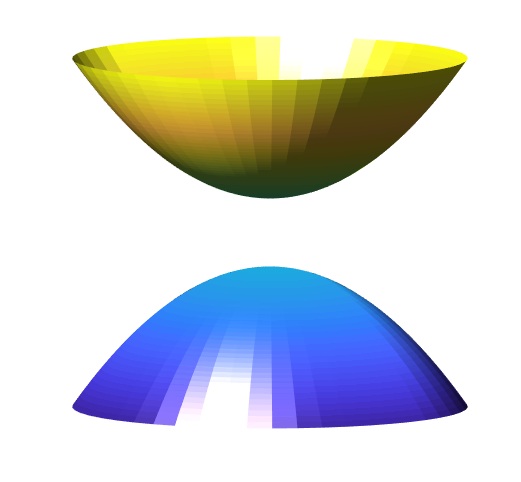}
    \end{minipage}
     \\ \hline
  \end{tabular}
\caption{Typical bipartite lattices include the square, honeycomb, and brickwall structures. The square lattice is a Bravais type with a single atom per unit cell, lacking chiral symmetry and fermion doubling. In contrast, honeycomb and brickwall lattices with two atoms per unit cell exhibit fermion doubling, although this feature may vanish, such as in the brickwall lattice when $t'/t>2$ \cite{Montambaux2009}. The Lieb lattice has a unit cell with 3 atoms, hence, it is bipartite but lacks fermion doubling.}\label{tbl:lattice_chiral_bipatite_valley}
\end{table}

\subsection{Valley degree of freedom - Dirac materials}

Bipartite lattices are associated with chiral symmetry (\ref{eq:chiral_symmetry}), a property which allows, under certain conditions, the emergence of new pseudo-spin degrees of freedom. It is possible to split the Hilbert space of free fermions on a non-Bravais bipartite lattice into the product $\mathcal{L}_{\text{latt}} \otimes \mathcal{S}^2 _{AB}$. Here $\mathcal{L}_{\text{latt}}$ accounts for orbital quantum states defined for the unit cells
and $\mathcal{S}^2 _{AB}$ is the Hilbert space of a two-state pseudo-spin that accounts for the two atoms $A$ and $B$, in the unit cell.
Examples of bipartite lattices are presented in fig.\ref{figure:symmetries in lattices}. 

For some chiral symmetric fermion Hamiltonians on non-Bravais bipartite lattices,
 an additional pseudo-spin degree of freedom emerges. It is associated to the presence of independent singular degenerate points $(K,K^{\prime})$ in the Brillouin zone where the energy gap vanishes. With linear energy dispersion near these points, the low-energy behavior is governed by a Dirac-like equation; hence they are termed Dirac points, Dirac cones, or independent valleys $(K,K^{\prime})$, as we will refer to them. Thus, the Hilbert space rewrites \begin{equation}
\tilde{\mathcal{L}}_{\text{latt}} \otimes \mathcal{S}^2_{KK'} \otimes \mathcal{S}^2 _{AB}.
\label{Hilbertspace}
\end{equation}

The pseudo-spin valley degree of freedom hinges on the closure of an energy gap in the Bloch Hamiltonian's spectrum. This leads to Dirac cones forming in pairs due to fermion doubling. This important and general result, encapsulated in the Nielsen-Ninomiya theorem \cite{Nielsen1983} states that massless and chiral Dirac fermions on a lattice necessitate additional fermion species.
This theorem applies to all lattices, not just bipartite ones. However, in condensed matter physics, the effective fermion field is a result of low energy electronic excitations. For which, effective chiral fermions require either a bipartite lattice or Sa Bogoliubov-de Gennes Hamiltonian for superconductors that satisfies (\ref{eq:chiral_symmetry}).

However, not all non-Bravais bipartite lattices with chiral symmetry show fermion doubling. For instance, the Lieb and brickwall $( t'/t > 2)$ lattices possess these properties yet they do not display fermion doubling. The Lieb lattice contains three atoms per unit cell, giving rise to a pseudospin $1$ rather than a pseudospin ${1}/{2}$. Its energy spectrum features a flat band intersecting the Dirac point, meaning all three bands must be taken into account. As a result, one cannot isolate a region where only two bands intersect to form an effective pseudospin ${1}/{2}$ model. In the case of the brickwall lattice with a Hamiltonian characterized by $ t'/t > 2$, the pseudospinor is massive, invalidating one of the fermion doubling requirements as displayed in Table \ref{tbl:lattice_chiral_bipatite_valley}. 

Fermion doubling and two valleys pseudo-spin are related to the assignment of an integer to each Dirac point, a property related to local Berry curvature.
In two dimensions, a general and elegant homotopy argument at the basis of the Nielsen and Ninomiya no-go theorem \cite{Nielsen1983} states that the sum of the winding numbers at the singular points is always 0. Generically (in the absence of lattice symmetries that would lead to a more specific behavior), the singular points all have winding numbers ±1, corresponding to gapless Weyl fermions of one chirality or the other. In this case, the vanishing of the sum of the winding numbers means that there are equally many gapless modes of positive or negative chirality, as expected for anomaly cancellation in a relativistic field theory. For condensed matter lattice Hamiltonians, the topological requirement about the existence of two valleys $(K,K')$ pseudo-spin and the Hilbert space (\ref{Hilbertspace}) is independent of perturbations as long as they do not break chirality nor the existence of a vanishing gap (massless fermions) in the energy spectrum \cite{Katsnelson2012}.

We focus now on bipartite lattices characterized by chiral Hamiltonians with paired valleys, which depend on lattice specifics and symmetries. These Hamiltonians, akin to chiral Dirac fermions, are represented by means of Clifford algebras \cite{Clifford1871, Dirac1928}, specifically involving anti-commuting Dirac matrices whose size and number depend on the number of two-state pseudo-spins \cite{Goft2023}. 
Thus, in the Hilbert space (\ref{Hilbertspace}), $2^2 \times 2^2$ matrices are needed to represent nearest-neighbors tight-binding Hamiltonians. This easily extends to $2^n \times 2^n$ matrices where $n$  represents the total number of involved pseudospin $1/2$ additional degrees of freedom such as layers or electronic spins. In the following, we examine spinless fermions on bipartite lattices with valleys $(K, K')$, defined within the effective Hilbert space (\ref{Hilbertspace}). This setup  
is defined by the four states: $\left( |AK \rangle, |BK \rangle, |AK' \rangle, |BK' \rangle \right)$, where $\left( |A \rangle, |B \rangle \right)$ are assigned to the bipartite lattice and $\left( |K \rangle, |K' \rangle \right)$ signify the two valleys, with the notation $|AK \rangle \equiv |A \rangle \otimes |K \rangle$. Hence, we respectively define  the lattice $\mathcal{B}_{\text{AB}}$ and the valley $\mathcal{B}_{\text{KK'}}$ bases by,
\begin{equation}
    \mathcal{B}_{\text{AB}} \equiv \left( |AK \rangle ,|AK' \rangle ,|BK \rangle, |BK' \rangle \right)
    \label{latticebasis}
\end{equation}
and
\begin{equation}
    \mathcal{B}_{\text{KK'}} \equiv \left( |AK \rangle ,|BK \rangle ,|AK' \rangle, |BK' \rangle \right) \, .
    \label{valleybasis}
\end{equation}
Setting the bases and matrix dimensions limits the number of independent anti-commuting matrices for the Clifford algebra to $5$. 
Since, as mentioned, these pseudo-spin degrees of freedom are independent of perturbations that do not break chirality and do not open an energy gap \cite{Slonczewski1958,Katsnelson2012} such as hopping disorder, the use 
of these two bases will still provide a good description in the presence of spatial defects breaking translation invariance. 

Effectively, the Hamiltonian (\ref{eq:ab_hamiltonian1}) can be written in the basis $\mathcal{B}_{\text{KK'}}$ 
by means of a low energy description restricted to first order in $a\delta k\ll1$ where $a$ is the lattice constant. This allows to block diagonalize (\ref{eq:ab_hamiltonian}), 
\begin{equation}\label{eq:valley_basis_Hamiltonian}
    \mathcal{H}=\psi^\dagger{H} \psi
\end{equation}
\begin{equation}
     {H} =\left(\begin{array}{cc}
{H}_{KK} & 0\\
0 & {H}_{K'K'}
\end{array}\right)
\label{h0valleys}
\end{equation}
where now $\psi=\left(\begin{array}{cccc}
a_{K} & b_{K} & a_{K'}& b_{K'}\end{array}\right)^T$  and the subset $\left( a_{K} , b_{K} \right)$ describes $N$ annihilation operators of sublattices $A$ and $B$ 
around valley $K$ (resp. $\left( a_{K'} , b_{K'} \right)$) and valley $K'$, and both ${H}_{KK}$ and ${H}_{K'K'}$ denote the Hamiltonian (\ref{eq:ab_hamiltonian}) expanded around each respective valley.

\begin{figure}[htbp]
    \centering
      {\includegraphics[width =0.5\columnwidth,bb=10bp 0bp 710bp 650bp,clip]{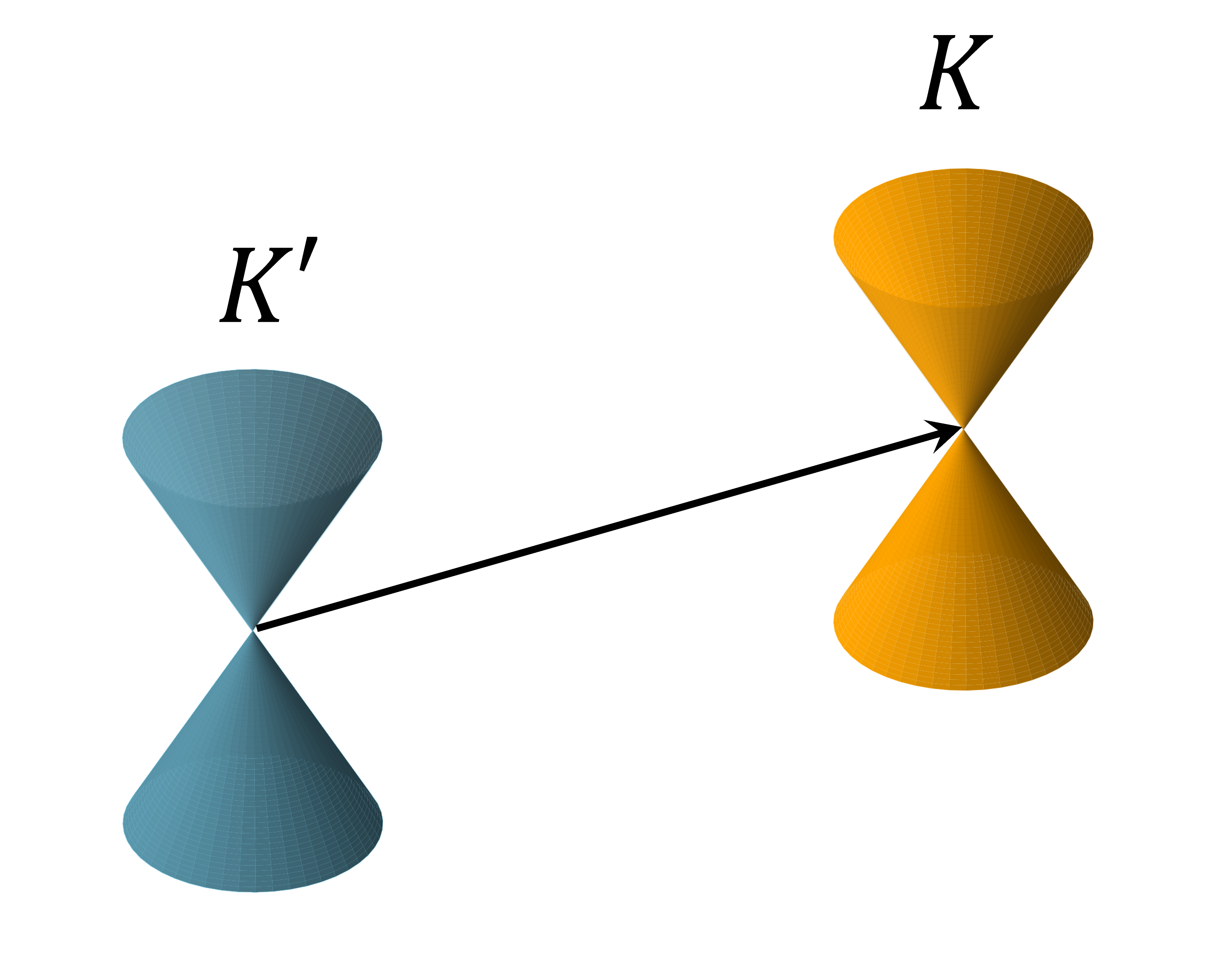}}
	\caption{Two Dirac cones $K$ and $K'$ display linear dispersion. Excitations of sublattices $A$ and $B$ arise around each cone, adding a pseudo-spin degree of freedom. The black arrow indicates the coupling between the cones due to local defects.}
\label{fig:two dirac cones}
\end{figure}

Local defects involve high momentum hence valley coupling terms, accounted for by a matrix  Hamiltonian with two types of $N \times N$ off-diagonal terms ${H}_{KK'}$ and ${H}_{K'K}$. This setup organizes operators by valley, independently of specific spatial or momentum representations, hence enabling a systematic framework to investigate the role of valley coupling and translational symmetry breaking defects in bipartite lattices as sketched in fig.\ref{fig:two dirac cones}. In the next section, we implement these considerations for the honeycomb lattice. 

%================ Figure 2 =================
%================ Ending Figure 2 ===========

\subsection{Honeycomb: Lattice and valley basis}

Two-dimensional bipartite Dirac materials, such as graphene modeled on a honeycomb lattice \cite{Novoselov2004}, possess unique properties \cite{Neto2009} that have led to a broad investigation for other similar materials. Alongside graphene, Dirac materials such as silicene and germanene \cite{Cahangirov2009,Wang2015} also exhibit a honeycomb lattice. Other materials like graphynes with different lattice arrangements have also been suggested \cite{Malko2012,Huang2013}.

\begin{figure}[t]
    \centering
      {\includegraphics[width =0.5\columnwidth,bb=200bp 130bp 505bp 388bp,clip]{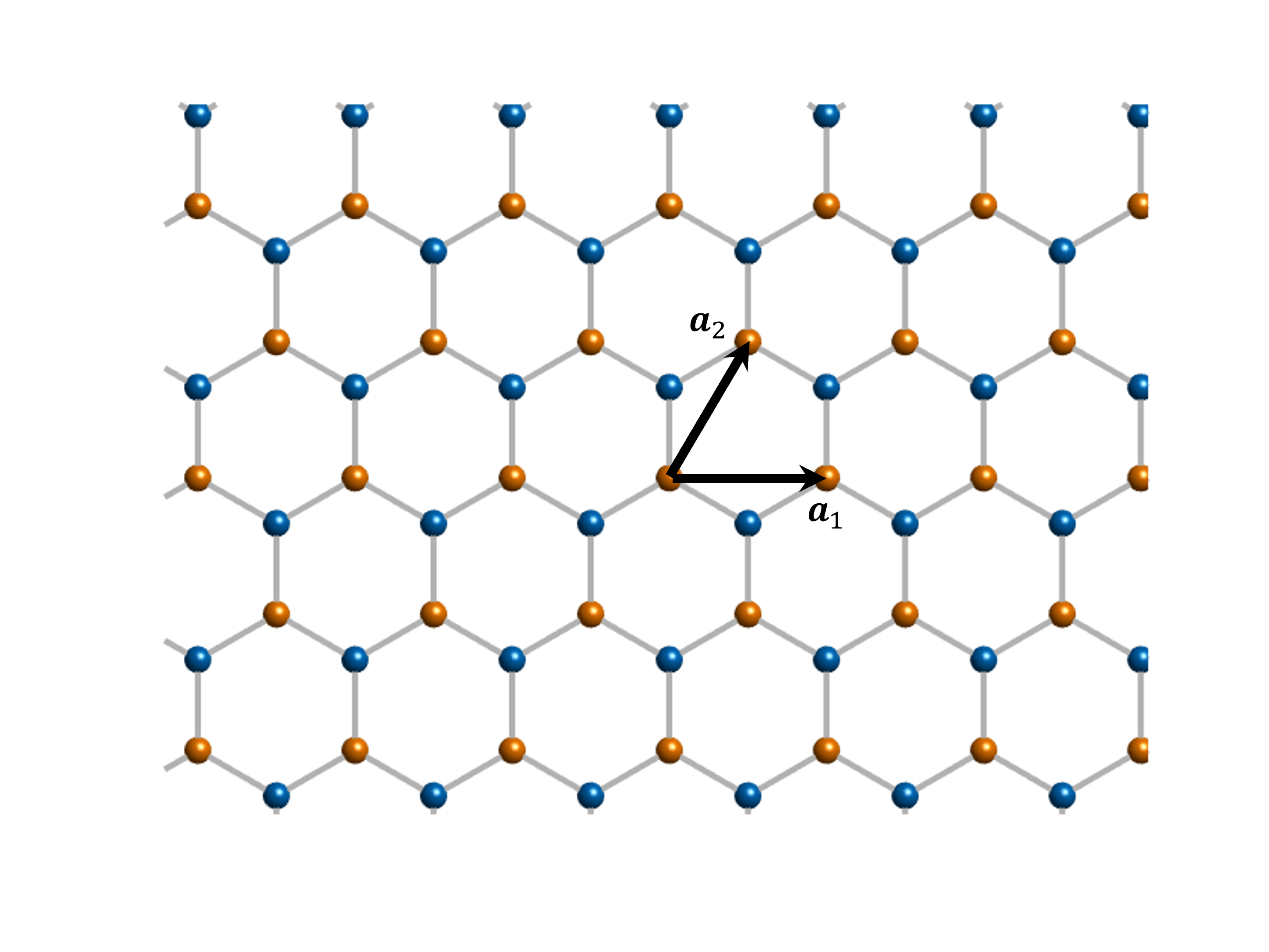}}
	\caption{Two triangular sublattices $A$ (orange) and $B$ (blue) form a honeycomb lattice, with lattice vectors shown.}
\label{fig:honeycomb lattice with vectors}
\end{figure}

The honeycomb lattice exemplifies a bipartite $(A,B)$ setup for applying the above methodology, with identical atoms at each site allowing on-site terms to be proportional to unity. Thus, the Hamiltonian of size $2N \times 2N$ in equation (\ref{eq:ab_hamiltonian}) is expressed as \cite{Wallace1947,Neto2009}: 
\begin{equation}\label{eq:tight_binding_H_0}
\mathcal{H}_{0}=-t\sum_{\substack{i=0,1,2\\ \boldsymbol{R}\in A}}a_{\boldsymbol{R}}^{\dagger}b_{\boldsymbol{R}+\boldsymbol{\delta}_{i}}+h.c. 
\end{equation}
where $t$ represents the hopping energy between nearest neighbors, and the constant onsite energy is set to zero. Using the lattice vectors (in terms of the lattice constant $a$) $\boldsymbol{a}_1$ and $\boldsymbol{a}_2$, given by 
$\boldsymbol{a}_{1}=\sqrt{3} a \, \boldsymbol{\hat{x}}$ and $\boldsymbol{a}_{2}=\frac{\sqrt{3}}{2} a \, \boldsymbol{\hat{x}}+\frac{3}{2}a \, \boldsymbol{\hat{y}} $, the vectors $\boldsymbol{\delta}_{i}$ joining nearest-neighbor sites are,  
$\boldsymbol{\delta}_{0}= \boldsymbol{0}$, $\boldsymbol{\delta}_{1}=-\boldsymbol{a}_{1}+\boldsymbol{a}_{2}$ and $\boldsymbol{\delta}_{2}=\boldsymbol{a}_{2}$ as represented in fig.\ref{fig:honeycomb lattice with vectors}. Chiral symmetry (\ref{eq:chiral_symmetry}) holds when ${H}_{aa}={H}_{bb}=0$, indicating no on-site or next-nearest neighbor terms in the Hamiltonian.\\
%================ Figure 3 =================
%================ Ending Figure 3 ===========
The creation operator $a_{\boldsymbol{R}}^{\dagger}$ is expressed via lattice momentum operators as $a_{\boldsymbol{R}}^{\dagger}=\frac{1}{\sqrt{N}}\sum_{\boldsymbol{k}}a_{\boldsymbol{k}}^{\dagger}e^{i\boldsymbol{k}\cdot\boldsymbol{R}}$, implying that,
\begin{align} \label{eq:H_sum_k}
\mathcal{H}_0=t\sum_{\boldsymbol{k}}f^{\vphantom{\ast}}\left(\boldsymbol{k}\right) a_{\boldsymbol{k}}^{\dagger}b_{\boldsymbol{k}}^{\vphantom{\ast}}+h.c.
\end{align}
where $f\left(\boldsymbol{k}\right) \equiv e^{-i\boldsymbol{k}\cdot\boldsymbol{\delta_{0}}}+e^{-i\boldsymbol{k}\cdot\boldsymbol{\delta_{1}}}+e^{-i\boldsymbol{k}\cdot \boldsymbol{\delta_{2}}}$. Recasting the Hamiltonian into Bloch form 
\begin{equation}
    \mathcal{H}_0=\int_{BZ} d\boldsymbol{k}\, \psi^{\dagger}_{\boldsymbol{k}} \, {H}_0 \, \psi^{\phantom{\ast}}_{\boldsymbol{k}} \, .
\label{HBZ} \end{equation}
The integration spans the first Brillouin zone, where $\psi_{\boldsymbol{k}} \equiv\left(\begin{array}{cc}\psi_{a,\boldsymbol{k}} & \psi_{b,\boldsymbol{k}}  \end{array}\right)^T$ represents the momentum basis sublattice spinor. Meanwhile, \begin{align}\label{eq:graphene_low_energy_hamiltonian} {H}_{0} \left(\boldsymbol{k}\right) & =t\left(\begin{array}{cc} 0 & f\left(\boldsymbol{k}\right)\\ f^*\left(\boldsymbol{k}\right) & 0 \end{array}\right) \end{align} is the $2 \times 2$ Bloch Hamiltonian. The energy spectrum  $\epsilon\left(\boldsymbol{k}\right)=\pm t\left|f\left(\boldsymbol{k}\right)\right|$ displays all the features described earlier for massless chiral fermions on a bipartite lattice.  It is gapped except for a pair of  independent Dirac points, $(\boldsymbol{K},\boldsymbol{K'})$ with  $\boldsymbol{K}= - \boldsymbol{K'} = \left(\frac{4\pi}{3\sqrt{3}a},0\right)$ 
at which $\epsilon\left(\boldsymbol{k}\right)$ vanishes with a linear dispersion at small $|\boldsymbol{k}|$ \cite{Wallace1947, Katsnelson2012,Neto2009}. 

The valley pseudo-spin degree of freedom is isolated by partitioning the momentum summation near the $\boldsymbol{K}$ and $\boldsymbol{K'}$ valleys, namely  $\sum_{\boldsymbol{k}}=\sum_{\boldsymbol{k}}^{\boldsymbol{K}}+\sum_{\boldsymbol{k}}^{\boldsymbol{K'}}$. Within each sum, the momentum is expanded around its specific valley, with the creation and annihilation operators modified accordingly \cite{Neto2009}. Around the $K$ valley, the momentum $\boldsymbol{k}$ is expressed as $\boldsymbol{k}=\boldsymbol{K}+\delta\boldsymbol{k}$, so that the annihilation operator for the A sublattice is redefined to be $a_{\delta\boldsymbol{k}}^{K}=e^{-i \boldsymbol{K} \cdot \boldsymbol{R}} \, a_{\boldsymbol{k}}$. In the continuum and low energy limit $a\delta k\ll1$, the function $f\left(\boldsymbol{K}+\delta\boldsymbol{k}\right) \simeq v_{F} \left( - \delta k_x - i \delta k_y \right)$ can be approximated to the first order in $a\delta k$ with  the Fermi velocity $v_{F} \equiv \frac{3}{2}ta$. This results in a $4 \times 4$ Bloch Hamiltonian which in the valley basis $\mathcal{B}_{\text{KK'}}$ in (\ref{valleybasis}), takes the form (\ref{h0valleys}) of an operator in a Clifford algebra. The corresponding low-energy Hamiltonian is:
\begin{align}
{H}_{0} & =v_{F}\left(\begin{array}{cccc}
0 & L & 0 & 0\\
L^{\dagger} & 0 & 0 & 0\\
0 & 0 & 0 & -L^{\dagger}\\
0 & 0 & -L & 0
\end{array}\right) \, ,
\label{h0valley}
\end{align}
and $L\equiv \delta k_x-i\delta k_y$. The valley matrices ${H}_{KK}$ and ${H}_{K'K'}$ take the explicit form,
\begin{equation}
{H}_{KK}=\left(\begin{array}{cc}
0 & L \\
L^{\dagger} & 0 \\
\end{array}\right),\:\:\:{H}_{K'K'}=\left(\begin{array}{cc}
0 & -L^\dagger \\
-L& 0 \\
\end{array}\right)
\end{equation}
whereas in the sublattice basis $\mathcal{B}_{\text{AB}}$ in (\ref{latticebasis}), the chiral symmetry becomes apparent, 
\begin{align}\label{eq:graphene_low_energy_hamiltonian_sublattice}
{H}_{0} & =v_{F}\left(\begin{array}{cccc}
0 & 0 & L & 0\\
0 & 0 & 0 & -L^{\dagger}\\
L^{\dagger} & 0 & 0 & 0\\
0 & -L & 0 & 0
\end{array}\right)
\end{align}
showing that ${H}_{aa}={H}_{bb}=0$ and such that the matrices ${H}_{ba}$ and ${H}_{ba}^\dagger$ are explicit,
\begin{equation}
{H}_{ba}=\left(\begin{array}{cc}
L^{\dagger} & 0 \\
0 & - L \\
\end{array}\right).
\label{h0lattice}
\end{equation}
The two operators (\ref{h0valley}) and (\ref{eq:graphene_low_energy_hamiltonian_sublattice}) are defined in the effective four-dimensional pseudo-spin Hilbert space $\mathcal{S}^2 _{KK'} \otimes \mathcal{S}^2 _{AB}$. They can also be written in a first quantization form within the Hilbert space $\mathcal{L}^2 \left( \mathbb{R}^2 \right) \otimes \mathcal{S}^2 _{KK'} \otimes \mathcal{S}^2 _{AB}$ where $\mathcal{L}^2 \left( \mathbb{R}^2 \right)$ is the Hilbert space of square integrable functions from $\mathbb{R}^2$ to $\mathbb{C}$ 
where, ${H}_0$ and $L$ turn into operators defined in that Hilbert space. Notably, ${L} = {q}_x - i {q}_y$ turns into the non-Hermitian momentum-like operator, 
\begin{equation} \label{Lop}
    \hat{L} = -i\partial_x-\partial_y \, ,
\end{equation} 
while the field operator $\psi$ becomes a square integrable wave function, preserving the original sublattice and valley pseudo-spin basis. This reformulation, replacing the initial $2N \times 2N$ tight-binding form of the Hamiltonian by a continuous first quantized form, allows for a continuum description of the honeycomb lattice in position space, which proves advantageous when addressing localized defects. The Hamiltonian becomes,  
\begin{equation}
    {\hat H}_{0}= \left(\begin{array}{cc}
0 & \hat D_0 ^\dagger \\
\hat D_0 & 0 \\
\end{array}\right) ,\:\:\:{\hat D_0}=v_{F}\left(\begin{array}{cc}
\hat L ^\dagger & 0 \\
0 & - \hat L \\
\end{array}\right)
\label{firstq}
\end{equation}
where the operator $\hat L$ is now defined in the infinite dimensional Hilbert space $\mathcal{L}^2 \left( \mathbb{R}^2 \right)$. As will be discussed in section \ref{sec:TopoZM}, the topological features of a lattice in the presence of defects or the absence thereof are encoded in the corresponding Clifford algebra \cite{Goft2023,Teo2010}.

\section{Spatially localized defect potentials}
\label{sec:potentials}

This section explicitly uses the previous description to outline a method for determining tight binding potentials associated with spatially localized defects. This facilitates the identification of symmetries that are either maintained or disrupted from the original perfect lattice model. The introduction of a spatially localized defect results in large momenta terms coupling the two valleys \cite{DutreixThesis,Ando1998}. 

The potential due to a defect localized at $\boldsymbol{R}_0$ is expressed via pairs of creation and annihilation operators, $a^{\dagger}_{\boldsymbol{R}_0}$ and $a_{\boldsymbol{R}_0 +\boldsymbol{\delta_i}}^{\phantom{\ast}}$. Expanding each operator in Fourier space enables us to separate the momenta around each valley \cite{Neto2009}. Consequently, the pair double summations over the two valleys divide into four terms,
\begin{equation}\label{eq:double_sum_four_valley_terms}
    \sum_{\boldsymbol{k}}\sum_{\boldsymbol{k'}}=\sum^K_{\boldsymbol{k}}\sum^K_{\boldsymbol{k'}}+\sum^{K'}_{\boldsymbol{k}}\sum^{K'}_{\boldsymbol{k'}}+\sum^K_{\boldsymbol{k}}\sum^{K'}_{\boldsymbol{k'}}+\sum^{K'}_{\boldsymbol{k}}\sum^K_{\boldsymbol{k'}}.
\end{equation}
The first two terms represent intravalley couplings, 
while the last two account for intervalley couplings. 
Momentum-space creation and annihilation operators are defined around each valley. Ultimately, summing over momenta yields $\psi_{\boldsymbol{r}}$, capturing the spatial behavior of the local defect. 

Two examples illustrate the application of this approach to local defects. The first one corresponds to an adatom, namely the addition of an atom on a site or a substitution by another atom \cite{Yndurain2014,Nanda2013,Dutreix2016, Dutreix2019}. The tight-binding and effective potential are well established, though a systematic method to obtain the effective potential has not been presented before. The second example corresponds to a vacancy defect, namely the removal of an atomic site and the corresponding bonds, for which the tight-binding potential is known \cite{Pereira2008,Palacios2008}, but the effective potential has not been obtained. Previous studies have used an effective potential that disagrees with experimental results \cite{Dutreix2013, Dutreix2016,Nanda2013, Bena2008}, providing two wavefront dislocations \cite{Nanda2012} instead of one \cite{Abulafia2023}, which our effective potential manages to describe.

\subsection{Potential of an adatom}
Adding or replacing an atom at a lattice site $\boldsymbol{R}_0$ alters the local chemical potential, a situation modeled by the perturbation potential of strength $V_0$,
\begin{equation}\label{eq:adatom_tight_binding}
   \mathcal{ V}_{\text{ad}} (\boldsymbol{R}_0)=V_0 \, a^{\dagger}_{\boldsymbol{R}_0}a_{\boldsymbol{R}_0}^{\phantom{\ast}} 
\end{equation}
of $
\mathcal{H}_{0}$ in (\ref{eq:tight_binding_H_0}) for an $A$ type adatom.
%================ Figure 4 =================
\begin{figure}[htbp]
    \centering
      {\includegraphics[width =0.5\columnwidth,bb=200bp 198bp 470bp 385bp,clip]{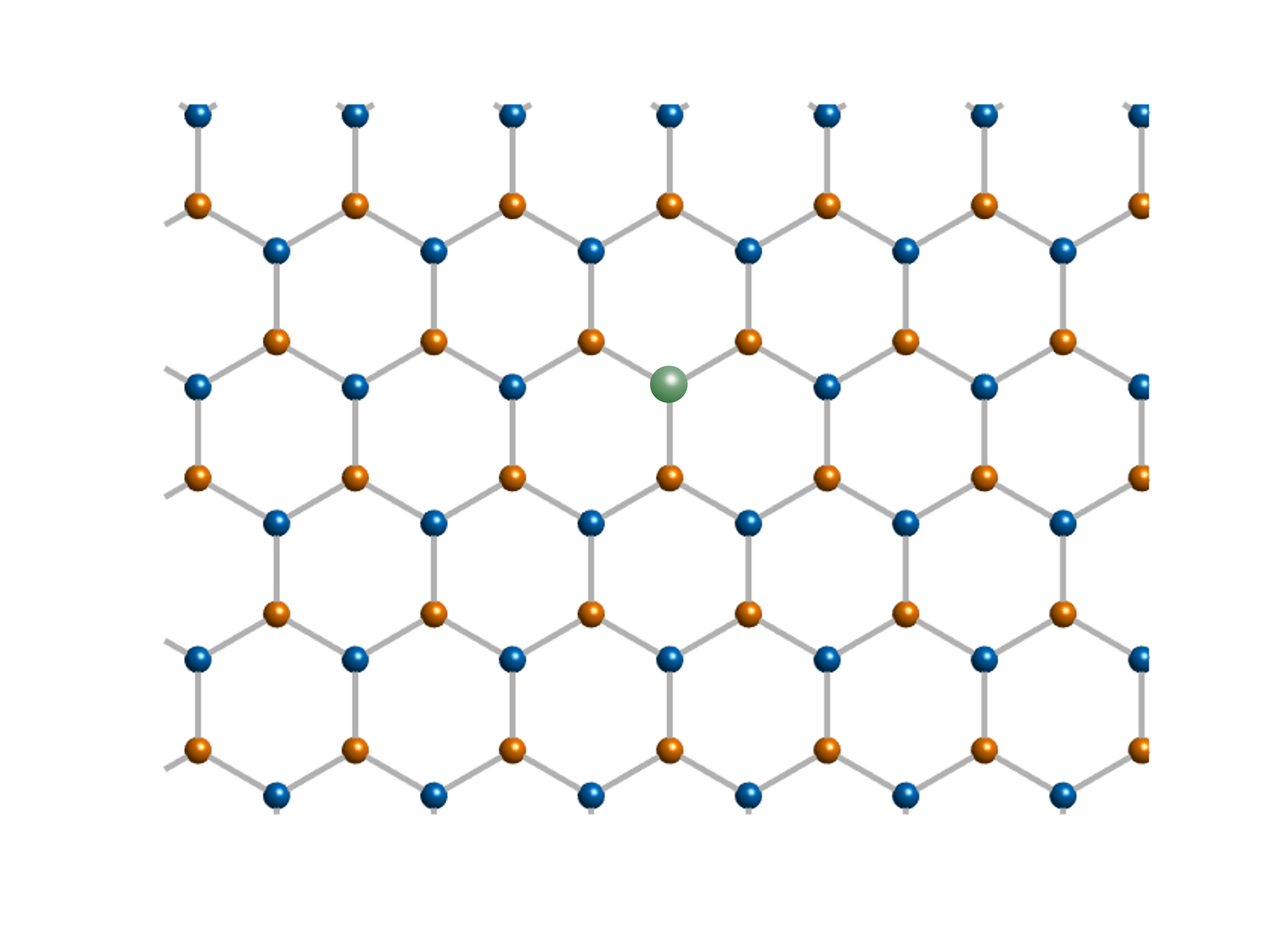}}
	\caption{Representation of an adatom potential in honeycomb lattice. 
    }
\label{fig:graphene with adatom}
\end{figure}
%================ Ending Figure 4 ===========

In the expression $\mathcal{V}_{\text{ad}} (\boldsymbol{R}_0)$, which follows (\ref{eq:ab_hamiltonian}), only the ${H}_{aa}$ block is non-zero, disrupting chiral symmetry. By converting each operator to a lattice momentum sum, we obtain
\begin{equation}
    \mathcal{V}_{\text{ad}} (\boldsymbol{R}_0)=\frac{V_0}{N}\sum_{\boldsymbol{k}}\sum_{\boldsymbol{k}'}e^{i\left(\boldsymbol{k}-\boldsymbol{k}'\right)\cdot \boldsymbol{R}_{0}}a_{\boldsymbol{k}}^{\dagger}a^{\vphantom{\ast}}_{\boldsymbol{k}'} \, .
\end{equation}
As specified by (\ref{eq:double_sum_four_valley_terms}), each sum divides into $K$ and $K'$ valleys, yielding the potential:
\begin{align}\label{eq:potential_operator_vallies_contributions}
\mathcal{V}_{\text{ad}} (\boldsymbol{R}_{0}) & =\mathcal{V}_{KK}+\mathcal{V}_{K'K'}+\mathcal{V}_{KK'}+\mathcal{V}_{K'K},
\end{align}
with  
\begin{align}
\mathcal{V}_{\alpha\beta}=\frac{V_{0}}{N}\sum_{\delta\boldsymbol{k}}^{\alpha}\sum_{\delta\boldsymbol{k}'}^{\beta}e^{i\left(\delta\boldsymbol{k}-\delta\boldsymbol{k}'\right)\cdot \boldsymbol{R}_{0}} \, {a_{\delta\boldsymbol{k}}^{\alpha}}^{\dagger}a_{\delta\boldsymbol{k}'}^{\beta} \, .
\end{align}
Defining $a_{\boldsymbol{R}_{0}}^{K}\equiv\frac{1}{\sqrt{N}}\sum_{\delta\boldsymbol{k}}^{K}e^{-i\delta\boldsymbol{k}\cdot\boldsymbol{R}_{0}}a_{\delta\boldsymbol{k}}^{K}$, leads to the effective adatom potential:
\begin{align}
\mathcal{V}_{\text{ad}}(\boldsymbol{R}_{0}) & =V_{0}\left(a_{\boldsymbol{R}_{0}}^{K}{}^{\dagger}a_{\boldsymbol{R}_{0}}^{K}+a_{\boldsymbol{R}_{0}}^{K'}{}^{\dagger}a_{\boldsymbol{R}_{0}}^{K'}\right)\\
 & +V_{0}\left(a_{\boldsymbol{R}_{0}}^{K}{}^{\dagger}a_{\boldsymbol{R}_{0}}^{K'}+a_{\boldsymbol{R}_{0}}^{K'}{}^{\dagger}a_{\boldsymbol{R}_{0}}^{K}\right).\nonumber
\end{align}

In first quantization, the operator ${\hat V}_{\text{ad}}(\boldsymbol{r})$ is derived by reconfiguring $\mathcal{V}_{\text{ad}}$ with spatial operators $\psi_{\boldsymbol{r}}$ using an integral in spatial form:
\begin{align}
\mathcal{V}_{\text{ad}}(\boldsymbol{R}_{0}) & =\int d\boldsymbol{r}\, \psi_{\boldsymbol{r}}^{\dagger}\, {\hat V}_{\text{ad}} (\boldsymbol{r})\, \psi^{\phantom{\ast}}_{\boldsymbol{r}}
\end{align}
and
\begin{equation}\label{eq:adatom_low_energy} 
{\hat V}_{\text{ad}}(\boldsymbol{r})= a^2 \, V_{0}\, \delta\left(\boldsymbol{r}-\boldsymbol{R}_{0}\right)\left(\begin{array}{cccc}
1 & 0 & 1 & 0\\
0 & 0 & 0 & 0\\
1 & 0 & 1 & 0\\
0 & 0 & 0 & 0
\end{array}\right) \, ,
\end{equation}
written in the valley basis $\mathcal{B}_{\text{KK'}}$ defined in (\ref{valleybasis}).  The operator $\delta\left(\boldsymbol{r}-\boldsymbol{R}_{0}\right)$ acts in the $\mathcal{L}^2 \left( \mathbb{R}^2 \right)$ Hilbert space and the $(4 \times 4)$ matrix in the four-dimensional pseudo-spin Hilbert space $\mathcal{S}^2 _{KK'} \otimes \mathcal{S}^2 _{AB}$. 
By representing (\ref{eq:adatom_low_energy}) within the lattice basis $\mathcal{B}_{\text{AB}}$, the potential naturally assumes the expected block diagonal structure,  
\begin{equation}
\label{eq:adatom_low_energy_sublattice} 
{\hat V}_{\text{ad}}(\boldsymbol{r})= a^2 \, V_{0} \, \delta\left(\boldsymbol{r}-\boldsymbol{R}_{0}\right)\left(\begin{array}{cccc}
1 & 1 & 0 & 0\\
1 & 1 & 0 & 0\\
0 & 0 & 0 & 0\\
0 & 0 & 0 & 0
\end{array}\right),
\end{equation}
given that only ${H}_{aa}$ is nonzero. Despite the availability of analytical solutions for the adatom potential \cite{Bena2008,Dutreix2021}, our method is highly effective for more intricate scenarios, such as the vacancy discussed next.
% --------------------------------------------------------------------------------
\begin{table*}
  \centering
  \begin{tabular}{ | m{4.8cm} | m{3.3cm} | m{4.7cm} | m{4.9cm} |  }
    \hline
    \centering    
      & \vspace{2pt} Adatom potential $\hat{V}_\text{ad}$  
    & Vacancy potential $\hat{V}_\text{vac}$   
     & Green's function $\hat{G}_{0,\epsilon}^R$  
         \\ \hline \footnotesize{$\mathcal{B}_{\text{AB}}=\left( |AK \rangle, |AK' \rangle ,|BK \rangle, |BK' \rangle 
         \right)$}
    & 
    \centering
        \vspace{0pt} \SmallMatrix{  \hat{V}_{AA}^{KK} & \hat{V}_{AA}^{KK'} &  & \\
        \hat{V}_{AA}^{K'K} & \hat{V}_{AA}^{K'K'} & \:\: \:\:&  \:\:\:\:\:\:\:\:\:\:\:\\
        &  & \:\: & \:\:\\
         & \: & \:\: & \:\:}
    &    \vspace{0pt}\footnotesize{$\left(\begin{array}{cccc}  &  &  & \hat{V}_{AB}^{KK'} \\
         &  & \hat{V}_{AB}^{K'K} &  \\
         & \hat{V}_{BA}^{KK'} &  & \\
        \hat{V}_{BA}^{K'K} &  &  & 
        \end{array}
        \right)$} 
    & 
\vspace{5pt}\footnotesize{$\left(\begin{array}{cccc} \hat{G}_{0,AA}^{K} &  & \hat{G}_{0,AB}^{K} &  \\
         & \hat{G}_{0,AA}^{K'} &  & \hat{G}_{0,AB}^{K'} \\
        \hat{G}_{0,BA}^{K} & & \hat{G}_{0,BB}^{K} &  \\
         &  \hat{G}_{0,BA}^{K'} &  & \hat{G}_{0,BB}^{K'} 
        \end{array}
        \right)$ }
        \\[22pt] \hline \footnotesize{$\mathcal{B}_{\text{KK'}} \equiv \left( |AK \rangle ,|BK \rangle ,|AK' \rangle, |BK' \rangle \right)$}
    &   
     \vspace{10pt}
       \footnotesize{ $\left(\begin{array}{cccc} \hat{V}_{AA}^{KK} &  \:& \hat{V}_{AA}^{KK'} &  \:\:\:\\
         & \: & \: & \:\\
        \hat{V}_{AA}^{K'K} & \: & \hat{V}_{AA}^{K'K'} &  \:\\
         & \: & \: & \:
        \end{array}
        \right)$}
    & 
    \vspace{5pt}
       \footnotesize{ $\left(\begin{array}{cccc}  &  &  & \hat{V}_{AB}^{KK'} \\
         &  & \hat{V}_{BA}^{KK'} &  \\
         & \hat{V}_{AB}^{K'K} &  & \\
        \hat{V}_{BA}^{K'K} &  &  & 
        \end{array}
        \right)$ }
    &
     \vspace{5pt}
       \footnotesize{ $\left(\begin{array}{cccc} \hat{G}_{0,AA}^{K} & \hat{G}_{0,AB}^{K} &  &  \\
        \hat{G}_{0,BA}^{K} & \hat{G}_{0,BB}^{K} &  &  \\
         &  & \hat{G}_{0,AA}^{K'} & \hat{G}_{0,AB}^{K'} \\
         &  & \hat{G}_{0,BA}^{K'} & \hat{G}_{0,BB}^{K'} 
        \end{array}
        \right)$ }
   \\[25pt]
    \hline
    \end{tabular}
	\caption{The adatom and vacancy potentials, along with the unperturbed retarded resolvent $\hat{G}_{0,\epsilon}^R $ for honeycomb lattice, are presented in two bases: lattice $\mathcal{B}_{\text{AB}}$ and valley $\mathcal{B}_{\text{KK'}}$, as specified in (\ref{latticebasis}) and (\ref{valleybasis}). 
 }
	\label{table:sublattice and valley basis}
\end{table*}
\subsection{Potential of a vacancy}
 A vacancy \cite{Kelly1998} is created by the removal of a neutral atom and its bonds. We examine vacancies that maintain rotational and chiral symmetries without forming new bonds \cite{Ugeda2010}. Since such vacancies only remove bonds, the potential for an $A$ type vacancy at site $\boldsymbol{R}_{0}$ is \cite{Pereira2006,Peres2009,Peres2009A}, 
\begin{equation}
\mathcal{V}_{\text{vac}} \left(\boldsymbol{R}_{0} \right) \equiv+ \, t\sum_{i=0,1,2}a_{\boldsymbol{R}_{0}}^{\dagger}b_{\boldsymbol{R}_{0}+\boldsymbol{\delta_{i}}}^{\vphantom{\ast}}+h.c. 
\label{VA}
\end{equation}
%================ Figure 5 =================
\begin{figure}[ht]
    \centering
      {\includegraphics[width =0.5\columnwidth,bb=200bp 198bp 470bp 385bp,clip]{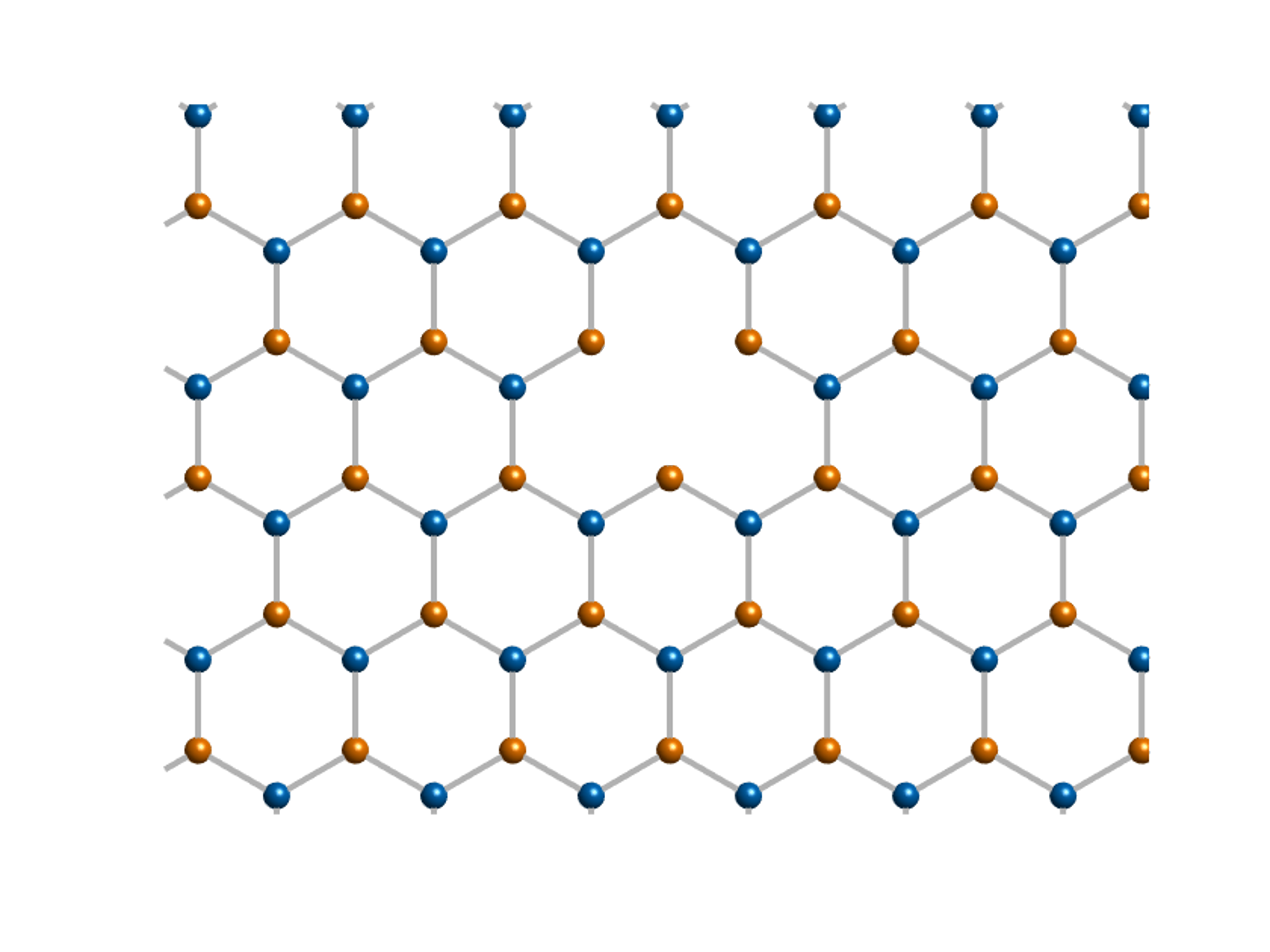}}
	\caption{A single carbon atom removed from a graphene honeycomb lattice, creating a vacancy. }
\label{fig:graphene with vacancy}
\end{figure}%================ Ending Figure 5 ===========
It only involves terms connecting sites $A$ to $B$, ensuring ${H}_{aa}={H}_{bb}=0$. Thus, the honeycomb lattice retains its initial chiral symmetry.
Deriving the first quantization expression of the potential for a vacancy resembles the process for an adatom, as it is a local spatial potential coupling valleys $K$ and $K'$. A Fourier decomposition of (\ref{VA}) leads to, 
\begin{align} \label{VAk}
\mathcal{V}_{\text{vac}}(\boldsymbol{R}_{0}) & =\frac{t}{N}\sum_{\boldsymbol{k},\boldsymbol{k'}}e^{i\left(\boldsymbol{k}-\boldsymbol{k}'\right)\cdot\boldsymbol{R}_{0}} \, f \left(\boldsymbol{k'} \right) \, a_{\boldsymbol{k}}^{\dagger}b_{\boldsymbol{k}'}^{\vphantom{\ast}}+h.c.
\end{align}
and $ f\left(\boldsymbol{k'} \right) =e^{-i\boldsymbol{k'}\cdot\boldsymbol{\delta_{0}}}+e^{-i\boldsymbol{k'}\cdot\boldsymbol{\delta_{1}}}+e^{-i\boldsymbol{k'}\cdot \boldsymbol{\delta_{2}}}$ accounts for missing bonds. The momentum sums split into four components, each representing coupling between sites $A$ and $B$, yielding eight distinct terms. For instance, the term,
\begin{align}
& \frac{1}{N}\sum_{\delta\boldsymbol{k}} ^{K} \sum_{\delta\boldsymbol{k}'} ^{K'}e^{i\left(\delta\boldsymbol{k}-\delta\boldsymbol{k}'\right)\cdot\boldsymbol{R}_{0}}\left(\delta k_{x}'+i\delta k_{y}'\right)a_{\delta\boldsymbol{k}}^{K\dagger}b_{\delta\boldsymbol{k}'}^{K'}
\end{align}
describes scattering from $B$ to $A$ sublattices and from valley $K'$ to $K$, with  
$ f\left(\boldsymbol{K}+\delta\boldsymbol{k}\right) \simeq v_{F} \left( - \delta k_x - i \delta k_y \right)$ to first order in $a\delta k \ll 1$. Summing over momenta $\delta\boldsymbol{k}$ and $\delta\boldsymbol{k}'$ leads to, 
\begin{align} a_{\boldsymbol{R}_{0}}^{K\dagger}\left(i\partial_{x_{0}}-\partial_{y_{0}}\right)b_{\boldsymbol{R}_{0}}^{K'} \, .
\end{align}
Among the eight terms in (\ref{VAk}), both intervalley and intravalley contributions are present. However, as will be discussed later, only the intervalley terms are topologically significant, contributing to zero-energy states and non-zero winding numbers \cite{Goft2023}. Therefore, we will concentrate on these four intervalley contributions:
\begin{align}\label{eq:vacancy_tight_binding_valley}
\mathcal{V}_{\text{vac}}(\boldsymbol{R}_{0}) & = v_F \, a_{\boldsymbol{R}_{0}}^{K\dagger}\left(i\partial_{x_{0}}-\partial_{y_{0}}\right)b_{\boldsymbol{R}_{0}}^{K'}+h.c.\nonumber \\
 & + v_F \, a_{\boldsymbol{R}_{0}}^{K'\dagger}\left(-i \, \partial_{x_{0}}-\partial_{y_{0}}\right)b_{\boldsymbol{R}_{0}}^{K}+h.c.
\end{align}
This second quantized form can be rewritten as
\begin{align}
 & \mathcal{V}_{\text{vac}}(\boldsymbol{R}_{0}) = v_F \int d\boldsymbol{r} \, a_{\boldsymbol{r}}^{K\dagger}\delta\left(\boldsymbol{r}-\boldsymbol{R}_0\right)\left(i \, \partial_{x}-\partial_{y}\right)b_{\boldsymbol{r}}^{K'}+h.c.\nonumber \\
 & + v_F \, \int d \boldsymbol{r} \,  a_{\boldsymbol{r}}^{K'\dagger}\delta\left(\boldsymbol{r}-\boldsymbol{R}_0\right)\left(-i \, \partial_{x}-\partial_{y}\right)b_{\boldsymbol{r}}^{K}+h.c.
\end{align}
so as to identify the first quantized potential ${\hat V}_{\text{vac}}$ from $\mathcal{V}_{\text{vac}}(\boldsymbol{R}_{0}) =  \int d^{2} \boldsymbol{r} \, \psi_{\boldsymbol{r}}^{\dagger} { V}_{\text{vac}} \, \psi_{\boldsymbol{r}}^{\vphantom{\ast}}$ written in the lattice basis $\mathcal{B}_{\text{AB}}$ defined in (\ref{latticebasis}),
\begin{equation}
{\hat V}_{\text{vac}} \equiv a^2v_F \left(
\begin{smallmatrix}
0 & 0 & 0 & -\delta\left(\boldsymbol{r-R_{0}}\right)\hat L^{\dagger}\\
0 & 0 & \delta\left(\boldsymbol{r-R_{0}}\right)\hat L & 0\\
0 & \hat L^{\dagger}\delta\left(\boldsymbol{r-R_{0}}\right) & 0 & 0\\
- \hat L \delta\left(\boldsymbol{r-R_{0}}\right) & 0 & 0 & 0
\end{smallmatrix}\right)\label{calV}
\end{equation}
where ${\hat L} $ defined in (\ref{Lop}). The operator ${\hat V}_{\text{vac}}$ acts in the $\mathcal{L}^2 \left( \mathbb{R}^2 \right)$ orbital Hilbert space and the $(4 \times 4)$ matrix in the pseudo-spin Hilbert space $\mathcal{S}^2 _{KK'} \otimes \mathcal{S}^2 _{AB}$. ${\hat V}_{\text{vac}}$ couples the valleys $K$ and $K'$ while maintaining chiral symmetry, unlike the adatom potential where chiral symmetry is broken \cite{Yndurain2014,Nanda2013,Dutreix2016}. Another essential feature of the vacancy potential appears when written in the lattice basis $\mathcal{B}_{\text{AB}}$. While it keeps the form (\ref{firstq}) of a chiral first quantized Hamiltonian, the addition to $\hat D_0$ of the operator ${\hat D_{\text{vac}}}$,
\begin{equation}
    {\hat D}_{\text{vac}}=a^2v_F\,\left(\begin{array}{cc}
0 & \hat L^{\dagger} \, \delta\left(\boldsymbol{r-R_{0}}\right) \\
 - \hat L \, \delta\left(\boldsymbol{r-R_{0}}\right) & 0 \\
\end{array}\right) 
\label{Dvacancy}
\end{equation}
changes the Hamiltonian into,
\begin{equation}
     {\hat H}_{\text{vac}}=\,\left(\begin{array}{cc}
0 & \hat{D}^{\dagger}  \\
 \hat{D} & 0 \\
\end{array}\right) 
\end{equation}
distinct from its counterpart in (\ref{firstq}). The operator $\hat D=\hat D_0+\hat D_{\text{vac}}$ possesses a non-zero analytical index, emerging since $\hat D \neq \hat D^\dagger$ is non-Hermitian. This leads to an angular momentum twist associated with the vacancy, producing a novel zero-energy state with a non-zero winding number. Further details are discussed in the next section.

\section{Topological Zero Modes}
\label{sec:TopoZM}

The Hamiltonian (\ref{eq:ab_hamiltonian}) characterizes non-Bravais bipartite lattices composed of two sublattices. The additional degrees of freedom give a pseudo-spin structure described by the Hilbert space $\mathcal{S}^2 _{KK'} \otimes \mathcal{S}^2 _{AB}$, leading to distinct topological features that we now explore. The bipartite nature is represented by the $2N \times 2N$ matrix in (\ref{eq:ab_hamiltonian}),
\begin{equation}
     H=\left(\begin{array}{cc}
H_{aa} & H_{ba}^\dagger\\
H_{ba} & H_{bb}
\end{array}\right) \, .
\label{Htopo}
\end{equation}
Chiral symmetry defined by (\ref{eq:chiral_symmetry}) implies that $H_{aa} = H_{bb} = 0$. Topological properties are encoded in the behavior of off-diagonal terms $H_{ba}$ and $H_{ba}^\dagger$. To retrieve these properties, our first quantization description proves helpful since it allows to compute topological invariants and to identify and characterize topological edge states. 

We consider the first quantized Hamiltonians $\hat H_0 + \hat V$, derived from (\ref{firstq}), (\ref{eq:adatom_low_energy_sublattice}), and (\ref{calV}) for adatom and vacancy potentials. 
The chiral Hamiltonian $\hat H_0 + \hat V_{\text{vac}}$ for an $A$ vacancy is of the form (\ref{firstq}) where $\hat D 
 \equiv \hat{D}_0+\hat{D}_{\text{vac}}$. An interesting class of quantum states called zero modes, are defined by $\hat D \, \psi_{A,\text{zm}}=0$. 
 
 The analytical index of an elliptic Dirac operator such as $\hat D$ is given by \cite{Nakahara1990, yankowsky2013},
\begin{equation}
    \text{Index } \hat D = \text{dim ker } \hat D -\text{dim ker } \hat D^\dagger \, .
    \label{indexanalyt}
\end{equation}
It captures topological properties by counting zero modes $\hat D \,\psi_{A,\text{zm}}=0$ on sublattice $A$ and correspondingly with $\hat D^\dagger $ on sublattice $B$. For elliptic operators defined on finite domains, this index is a finite integer. Clearly, the analytical index vanishes for a Hermitian operator $\hat D$. According to the Atiyah-Singer index theorem \cite{Atiyah1963, Atiyah1968,Niemi&Semenoff1984,yankowsky2013}, the analytical index is also a topological integer $\nu$ that can be computed through other means \cite{Goft2023}. It is easy to show that these zero modes are also zero energy eigenstates of the chiral Hamiltonian \cite{Stone1984}, namely $\left( \hat H_0 + \hat V_{\text{vac}} \right) \, \psi_{\text{zm}}=0$ and have a non-vanishing wavefunction on one sublattice only, either $A$ or $B$. We denote $N_{\text{zm}}$ the number of such zero modes enumerated by the index theorem. They are of topological origin, and they are related to the topological integer $\nu$ according to,
\begin{equation}
    |\text{Index } \hat D | = | \nu | = N_{\text{zm}} \, .
    \label{indexthm}
\end{equation}
This is an example of bulk-edge correspondence \cite{Hatsugai1993} which describes how zero modes $\psi_{\text{zm}}$, typically found on an edge, relate to a topological integer $\nu$ derived for an infinite system without boundaries.

Here, use has been made of the existence of chiral symmetry. Let us exploit it further before turning to other situations. The operator
$\hat D$ may be hermitian, in which case its analytical index identically vanishes since both $\text{ ker } \hat D$ and $\text{ker } \hat D^\dagger$ are identical and of finite dimensions. Hence, a necessary condition for having topological zero modes is for $\hat D$ to be non-hermitian. Another interesting situation occurs when, while being non-hermitian, both $\hat D$ and $\hat D^\dagger$ have the same number of zero modes, namely there is an equal number of zero modes spatially localized on sublattices $A$ and $B$. Hence, while being topologically non-trivial, such a system has a vanishing topological number.

Chiral symmetry can also be broken by adding a mass term $M$, namely for a Hamiltonian,
\begin{equation}
     \hat H=\left(\begin{array}{cc}
M & \hat D^\dagger\\
\hat D & - M
\end{array}\right) \, . 
\end{equation}
Clearly, the additional mass term is proportional to $\sigma_3$ so that (\ref{eq:chiral_symmetry}) does not hold any longer. Yet, the analytical index is still defined \cite{Goft2023} and the index theorem still reveals underlying topology. The only difference being that the mass term shifts the zero modes $\psi_{\text{zm}}$ to non-zero energies but without modifying the bulk-edge correspondence nor the topological nature of these modes. 

The next subsections propose an analysis of the potentials created by local defects as presented in section \ref{sec:potentials} in the light of topological features.

\subsection{Zero modes and adatoms}

The first quantized Hamiltonian of graphene with an adatom, equivalent to (\ref{Htopo}) and corresponding to the potential ${\hat V}_\text{ad} (\boldsymbol{r})$ induced by an adatom localized in $\boldsymbol{R}_0$ as given by (\ref{eq:adatom_low_energy_sublattice}) is, 
\begin{equation}
\hat H_{\text{ad}}={\hat H}_0+{\hat V}_\text{ad} (\boldsymbol{r}) \, ,
\end{equation}
where ${\hat H}_0$ given in (\ref{firstq}) accounts for pristine graphene. Since the adatom potential (\ref{eq:adatom_low_energy_sublattice}) has vanishing off-diagonal terms, the operators 
$H_{ba}$ and $H_{ba}^\dagger$ in $H_{\text{ad}}$ remain identical to those of pristine graphene. Both infinite pristine graphene and graphene with adatoms lack zero modes, rendering them non-topological and giving them a zero analytical index \cite{Ovdat2020}. Thus, although adatoms in graphene connect valleys $K$ and $K'$, they do not generate topological states.
 
 Graphene flakes of finite size can exhibit zero energy states by selecting particular boundary conditions, even when adatoms are present \cite{Weik2016}. These supernumerary zero modes appear on both sublattices $A$ and $B$, but they do not solve $D\psi_{A,\text{zm}}=0$. Consequently, they are neither countable via the analytical index (\ref{indexanalyt}) nor topological.

\subsection{Topological zero modes associated to vacancies}
The first quantized Hamiltonian corresponding to the potential created by a vacancy in pristine graphene is,
\begin{equation}
\hat H_{\text{vac}}={\hat H}_0+{{\hat V}_{\text{vac}}} = \left(\begin{array}{cc}
0 & \hat D_0^\dagger + \hat D_{\text{vac}}^\dagger \\
\hat D_0 + \hat D_{\text{vac}}& 0
\end{array}\right)
\label{vac}
\end{equation}
where ${\hat H}_0$ and $\hat D_0$ given in (\ref{firstq}) account for pristine graphene,  and ${{\hat V}_{\text{vac}}}$ and $\hat D_{\text{vac}}$ are respectively given by (\ref{calV}) and (\ref{Dvacancy}).
Since $\hat{L} \equiv -i\partial_x-\partial_y$ does not commute with $\delta\left(\boldsymbol{r}-\boldsymbol{R}_0\right)$, then $\hat D_{\text{vac}} \neq \hat D_{\text{vac}}^\dagger$, hence opening the possibility for a non-vanishing analytical index and topological zero modes determined by solutions to  
$(\hat D_0 + \hat D_{\text{vac}} )\, \psi_{A,\text{zm}}=0$ and $(\hat D_0^\dagger + \hat D_{\text{vac}}^\dagger ) \, \psi_{B,\text{zm}}=0$. We note that $\delta\left(\boldsymbol{r}-\boldsymbol{R}_0\right)$ is an approximation for the potential created by a vacancy. It can be replaced by a real and spatially localized function that vanishes for $r > r_0$, whose exact shape is not important.

For a vacancy at $\boldsymbol{R}_{0} = \boldsymbol{0}$, zero mode solutions $\psi_{\text{zm}}$ are sought in a scattering form for $r \gg r_0$ \cite{Ovdat2020}: 
\begin{equation}\label{eq:zm_ang_momentum}
    \left\langle \boldsymbol{r}|\psi_{\text{zm}}\right\rangle=\psi_{\text{zm}} \left(\boldsymbol{r}\right) = \sum_{m=-\infty}^{\infty} e^{im\theta}\left(\begin{array}{c}\psi^K_{A,m}\left(r\right)\\ \psi^{K'}_{A,m}\left(r\right)\\\psi^K_{B,m}\left(r\right) \\\psi^{K'}_{B,m}\left(r\right)\end{array}\right),
\end{equation}
using angular momentum expansion. Angular momentum in two dimensions is given by $\hat{L}_{\text{m}} \equiv -i(x \, \partial_y - y \, \partial_x)$, satisfying $\hat{L}_{\text{m}}\left|m\right\rangle= m\left|m\right\rangle$, where $\left\langle\boldsymbol{r}|m\right\rangle\propto e^{im\theta}$. It commutes with $\hat{L}$ and $\hat{L}^{\dagger}$ in (\ref{Lop}) akin to a ladder structure,
\begin{align}\label{eq:ladder_op_def}
    \hat{L}\left|m\right\rangle&=\left|m-1\right\rangle\\
    \hat{L}^{\dagger}\left|m\right\rangle&=\left|m+1\right\rangle \, . \nonumber
\end{align}

Zero modes are found by solving the Dirac Hamiltonian (\ref{firstq}) in lattice basis $\mathcal{B}_{\text{AB}}$, specifically $\psi_{A,m}^{K}\left(r\right)$ and $\psi_{B,m}^{K'}\left(r\right)$, with obvious notation, behaving as $r^m$, while $\psi_{A,m}^{K'}\left(r\right)$ and $\psi_{B,m}^{K}\left(r\right)$ exhibit $r^{-m}$. Requiring vanishing wavefunctions at infinity limits angular momentum $m$ as shown in Table \ref{Table:modes}. The remaining non-vanishing modes are given by:
\begin{align}\label{eq:vanishing_at_infty_modes}
\psi_{A,m}^{K}\left(r\right)&=A_{m}^{K}r^{m} ,\:\:\:\:m=-\infty,...,-1 \nonumber \\  \psi_{B,m}^{K}\left(r\right)&=B_{m}^{K}r^{-m},\:\:\:\:m=1,...,\infty\\
\psi_{A,m}^{K'}\left(r\right)&=A_{m}^{K'}r^{-m} ,\:\:\:\:m=1,...,\infty& \nonumber \\ \psi_{B,m}^{K'}\left(r\right)&=B_{m}^{K'}r^{m},\:\:\:\:m=-\infty,...,-1\nonumber
\end{align}
We determine zero mode solutions for sublattice $A$ by solving $\left\langle r, m' \right|(\hat{D}_0+\hat{D}_{\text{vac}})\left|\psi_{A,\text{zm}}\right\rangle=0$; this involves using (\ref{eq:zm_ang_momentum}) to find states $\left|m\right>$ with non-zero coefficients $A^{K,K'}_{m}$, as specified in \eqref{eq:vanishing_at_infty_modes}, 
\begin{align} 
\label{eq:zero_mode_equation_sublattice_A}
& \left\langle m'-1\right|\sum_{m=-\infty}^{-1}A_{m}^{K}r^{m}+a^{2}\sum_{m=1}^{\infty}A_{m}^{K'}\delta\left(\boldsymbol{r}\right)r^{-m}\left|m\right>=0\nonumber\\
 & \left\langle m'+1\right|\sum_{m=1}^{\infty}A_{m}^{K'}r^{-m}+a^{2}\sum_{m=-\infty}^{-1}A_{m}^{K}\delta\left(\boldsymbol{r}\right)r^{m}\left|m\right>=0.
\end{align}
In every equation, the operator $\hat{D}_0+\hat{D}_{\text{vac}}$ shares the same angular momentum operator, and using the ladder structure \eqref{eq:ladder_op_def} on $\left<m'\right|$ yields distinct equations for $A^{K}_{m}$ and $A^{K'}_{m}$ across all $m$, causing these coefficients to vanish, thus eliminating zero-mode solutions for sublattice A (see Table \ref{Table:modes}).

In contrast, sublattice $B$ involves distinct ladder operators in $\left\langle r,m'\right|(\hat D_0^\dagger + \hat D_{\text{vac}}^\dagger ) \,\left| \psi_{B,\text{zm}}\right\rangle =0$, linking angular modes $B_{1}^K$ and $B_{-1}^{K'}$,
\begin{align}\label{eq:zero_mode_equation_sublattice_B} & \left\langle m'\right|\sum_{m=1}^{\infty}B_{m}^{K}\hat{L}r^{-m}-a^{2}\delta\left(\boldsymbol{r}\right)\sum_{m=-\infty}^{-1}B_{m}^{K'}\hat{L}^{\dagger}r^{m}\left|m\right>=0\nonumber\\
 & \left\langle m'\right|\sum_{m=-\infty}^{-1}B_{m}^{K'}\hat{L}^{\dagger}r^{m}-a^{2}\delta\left(\boldsymbol{r}\right)\sum_{m=1}^{\infty}B_{m}^{K}\hat{L}r^{-m}\left|m\right>=0.
\end{align}
Therefore, sublattice $B$ admits a single non-vanishing mode as shown in Table \ref{Table:modes}. Equations (\ref{eq:zero_mode_equation_sublattice_B}) for sublattice $B$ remain unchanged when swapping valleys $K$ and $K'$, and with $\delta \left(\boldsymbol{r}\right)$ being real, the valley coefficients of the pseudospinor must be identical. Consequently, there is only one non-normalizable zero mode solution with a power law decay \cite{Pereira2006, Pereira2008}, expressed in the sublattice basis $\mathcal{B}_{\text{AB}}$ (\ref{latticebasis}) as,
\begin{table}
\small
\begin{center}\begin{tabular}{|c|c|c|c|c|}
\hline 
$m$ & $A_m ^K$ & $A_m ^{K'}$ & $B_m ^{K}$ & $B_m ^{K'}$ 
\\
\hline 
\hline 
 $\vdots$ & 0 & \cellcolor{lightgray} & \cellcolor{lightgray} &0\\
 \hline
3 & 0 & \cellcolor{lightgray} & \cellcolor{lightgray} & 0
\\
\hline 
2 & 0 & \cellcolor{lightgray} & \cellcolor{lightgray} & 0
\\ 
\hline
1 & 0 & \cellcolor{lightgray} & \cellcolor{orange} & 0
\\
\hline 
0 & 0 & 0 & 0 & 0
\\
\hline 
 -1 & \cellcolor{lightgray} & 0 & 0 & \cellcolor{orange}
\\
\hline 
  -2 & \cellcolor{lightgray} & 0 & 0& \cellcolor{lightgray}
\\
\hline 
 -3& \cellcolor{lightgray} & 0 & 0 & \cellcolor{lightgray}
\\
\hline 
 $\vdots$ & \cellcolor{lightgray} & 0 & 0 &\cellcolor{lightgray}

\\
\hline
\end{tabular}
\end{center}
\caption{Angular coefficients for each sublattice and valley are provided. The left column shows $m$ values, with zero denoting coefficients that vanish as $r\to\infty$. Only $ B_1 ^{K}$ and $ B_{-1} ^{K'}$ (highlighted in orange) are linked by the vacancy potential, resulting in a single zero-energy mode on sublattice $B$. All other coefficients (in grey) are zero and unaltered by the vacancy potential.
\label{Table:modes}}
\end{table}
\begin{equation}
\psi_{\text{zm}}\left(\boldsymbol{r} \right) =\left(\begin{array}{cc}  0 \, \, \,   \psi_{B,\text{zm}} \end{array}\right)^T = \frac{1}{r}\left(\begin{array}{c}
0\\
0\\
e^{i\theta}\\
e^{-i\theta}
\end{array}\right) \, .
\end{equation}

We now focus on the topological features of the zero mode $\psi_{\text{zm}}\left(\boldsymbol{r} \right)$, also a zero-energy eigenstate satisfying $\hat H_{\text{vac}} \, \psi_{\text{zm}}\left(\boldsymbol{r} \right) =0$. Since $\hat D_{\text{vac}} \neq \hat D_{\text{vac}}^\dagger$, the analytical index $\text{Index } \hat D = -1$ from (\ref{indexanalyt}) discloses the existence of one localized topological zero mode at the vacancy. The zero mode's spatial location is influenced by the choice of boundary conditions. This represents the "edge" in bulk-edge correspondence. A bulk topological invariant calculation is needed for the "bulk," as elaborated in \cite{Goft2023}, complying with (\ref{indexthm}). The bulk topological invariant is physically significant as a winding number $\nu$ that satisfies (\ref{indexthm}) and shows up as a dislocation pattern in electronic density \cite{Abulafia2023}.

 To investigate the bulk-edge correspondence  and potential-induced topological states as discussed in this section, we now assess the local electronic density calculated in the framework of Green's functions formalism.

\section{\label{sec:level4} Detecting Topological states using Green's Functions}
In section \ref{sec:TopoZM}, we studied the existence and properties of topological zero modes associated with a local potential. In this section, we demonstrate how such zero
modes quantitatively impact measurable properties expressed using Green's functions, like local electronic density or transport quantities. A key advantage of our first quantized method is its simplicity in computing Green's function variations for spatial behavior, intervalley coupling, or interactions among isospin degrees of freedom within the Hilbert space $\mathcal{L}^2 \left( \mathbb{R}^2 \right) \otimes \mathcal{S}^2 _{KK'} \otimes \mathcal{S}^2 _{AB}$. 

\subsection{\texorpdfstring{Green's functions in lattice $\mathcal{B}_{\text{AB}}$ and valley $\mathcal{B}_{\text{KK'}}$ bases}{Green's functions in lattice BAB and valley BKK' bases}}

The resolvent operator $\hat{G}_{\epsilon}$ at energy $\epsilon$, associated with the first quantized Hamiltonian ${\hat H} = {\hat H}_0 + {\hat V}$ is defined by  $\hat{G}_{\epsilon}  \equiv ({\epsilon} \mathbf{1} - {\hat H} )^{-1} $. The operator  ${\hat V} $ represents the localized potential associated either with an adatom or with a vacancy. A standard way to compute the resolvent starts from the series expansion in terms of the potential,
\begin{equation}
     \delta \hat{G}_{\epsilon}  \equiv \hat{G}_{\epsilon} -\hat{G}_{0,{\epsilon} } = \hat{G}_{0,\epsilon}\hat{V}\hat{G}_{0,\epsilon}+\hat{G}_{0,\epsilon}\hat{V}\hat{G}_{0,\epsilon}\hat{V}\hat{G}_{0,\epsilon}+...
     \label{series}
 \end{equation}
 where $\hat{G}_{0,\epsilon}$ is the resolvent associated with the unperturbed Hamiltonian $\hat H_0$. These resolvent operators are defined in the total Hilbert space $\mathcal{L}^2 \left( \mathbb{R}^2 \right) \otimes \mathcal{S}^2 _{KK'} \otimes \mathcal{S}^2 _{AB}$. To obtain relevant physical quantities in perturbation theory, we use standard expansion methods in powers of the potential, together with partial truncation of the self-energy and a convenient Dyson series resummation of irreducible diagrammatic contributions \cite{Akkermans2007}. The isospin degrees of freedom associated with the bipartite lattice and valleys, characterized by $\mathcal{S}^2 _{KK'} \otimes \mathcal{S}^2 _{AB}$, become relevant in the presence of defects, enabling new valley couplings and yielding topological zero modes.
 
We introduce some useful quantities. The Green's function's spatial behavior is expressed by the matrix element $\langle \boldsymbol{r} | \delta \hat{G}_{\epsilon} | \boldsymbol{r'} \rangle $ in position representation $|\boldsymbol{r} \rangle $ within $\mathcal{L}^2 \left( \mathbb{R}^2 \right)$. This quantity is a $4 \times 4$ matrix defined in $\mathcal{S}^2 _{KK'} \otimes \mathcal{S}^2 _{AB}$. 
Valley coupling terms are defined as well. Specifically, $\delta \hat{G}_{\epsilon} ^{KK'} \equiv \langle K' | \delta \hat{G}_{\epsilon} | K \rangle$ represents a $2\times2$ matrix  within $\mathcal{L}^2 \left( \mathbb{R}^2 \right)\otimes \mathcal{S}^2 _{AB}$. 
Next, we define the trace operation on the two-state pseudo-spin $\mathcal{S}^2_{AB}$ for the bipartite lattice, acting on the sublattice $(A,B)$ degrees of freedom as follows:
\begin{equation}\label{eq:greens_function_all_valley_contribution2}
\text{Tr}_{AB}\, \delta \hat{G}_{\epsilon}^{KK'} \equiv\left\langle A\right|\delta \hat{G}_{\epsilon}^{KK'}\left|A\right \rangle +\left\langle B\right|\delta \hat{G}_{\epsilon}^{KK'}\left|B\right \rangle 
 \end{equation}
 where each term is an operator defined in $\mathcal{L}^2 \left( \mathbb{R}^2 \right) $.

For the unperturbed Hamiltonian ${\hat H}_0$ in (\ref{firstq}), where valleys and lattice degrees of freedom remain separate \cite{Bena2008,Neto2009}, these quantities are straightforward to calculate. The free Green's function $\hat G_{0,\epsilon}\left(\boldsymbol{r}\right)$, defined as $\mbox{diag} \{ \hat G_{0,\epsilon}^{K} (\boldsymbol{r}), \hat G_{0,\epsilon}^{K'} (\boldsymbol{r})\}$, is expressed within the valley basis $\mathcal{B}_{\text{KK'}}$. The valley Green's functions, $\hat G_{0,\epsilon}^K$ and $\hat G_{0,\epsilon}^{K'}$ are $2\times2$ matrices defined in $\mathcal{S}^2 _{AB}$, 
by \cite{Bena2008, Dutreix2016}, 
\begin{align}\label{eq:graphene Green function valley_K}
\hat G_{0,\epsilon}^{K }\left(\boldsymbol{r}\right)& = e^{i\boldsymbol{K }\cdot\boldsymbol{r}}\left(\begin{array}{cc}
\frac{\epsilon}{v_{F}}{g} (\epsilon \, r) & \hat L \, {g} (\epsilon \, r)\\
\hat L^\dagger{g} (\epsilon \, r) & \frac{\epsilon}{v_{F}}{g} (\epsilon \, r)
\end{array}\right)\\  \nonumber
\hat G_{0,\epsilon}^{K'}\left(\boldsymbol{r}\right)& = e^{i\boldsymbol{K}'\cdot\boldsymbol{r}}\left(\begin{array}{cc}
\frac{\epsilon}{v_{F}}{g} (\epsilon \, r) & -\hat L^\dagger{g} (\epsilon \, r)\\
- \hat L \,{g} (\epsilon \, r) & \frac{\epsilon}{v_{F}}{g} (\epsilon \, r)
\end{array}\right)
\end{align}
where ${g} (\epsilon \, r) \equiv \frac{- i}{4 v_{F}}H_{0}^{(1)} (\epsilon \, r / v_{F})$ is defined using a Hankel function, and $\hat L$ from (\ref{Lop}) which follows, 
\begin{equation}\label{Lang}
\begin{cases}
\hat L{g} (\epsilon \, r)=-ie^{-i\theta}\frac{d}{dr}{g} (\epsilon \, r)\\
\hat L^{\dagger}{g} (\epsilon \, r)=-ie^{i\theta}\frac{d}{dr}{g} (\epsilon \, r) \, .
\end{cases}
\end{equation}
plays an important role in topological valley coupling.

\subsection{Valley coupling electronic density}

A local potential causes valley coupling characterized by ${\Delta K} \equiv \boldsymbol{K} -\boldsymbol{K'}= \frac{8\pi}{3\sqrt{3}a}$, resulting in spatial oscillations with a wavelength $2\pi/\Delta K = 3 \sqrt 3 a /4$, thereby altering the local electronic density $\delta\rho_{\Delta K} \left(\boldsymbol{r}\right)$, which is influenced by this coupling.

In a bipartite lattice $(AB)$ with valleys $(KK')$, the change of local electronic density due to  valley coupling is given by \cite{Akkermans2007,Dutreix2019},
\begin{equation}\label{eq:local density and local density of states}
  \delta\rho_{\Delta K} \left(\boldsymbol{r}\right)= - \int  \frac{d\epsilon}{\pi} \, \text{Im }\left\langle\boldsymbol{r}\right| \text{Tr}_{AB} \delta \hat{G}_{\epsilon_+} ^{KK'} \left|\boldsymbol{r}\right\rangle + (K \leftrightarrow K' )  
\end{equation}
where the retarded resolvent is defined by choosing the complex-valued energy $\epsilon_{+} \equiv \epsilon + i0^+$.  
Using time-reversal symmetry, which swaps $K\leftrightarrow K'$ and is analogous to complex conjugation,
\begin{align}\label{K'K}
\delta G_{\epsilon_+}^{K'K}\left(\boldsymbol{r},\boldsymbol{r}\right) & = \left( \delta G_{\epsilon_-}^{KK'} \left(\boldsymbol{r},\boldsymbol{r}\right) \right)^{*} \, ,
\end{align}
 allows computing $\langle \boldsymbol{r} |\delta G_{\epsilon_+}^{KK'}| \boldsymbol{r} \rangle$ only to obtain  $\delta\rho_{\Delta K} \left(\boldsymbol{r}\right)$. 

Calculating just the initial term $\delta \hat G_{\epsilon_+} ^{ (1)} = \hat{G}_{0,\epsilon_+}\hat{V}\hat{G}_{0,\epsilon_+}$ from the series expansion (\ref{series}) suffices to capture key topological characteristics. The matrix element $\delta G_{\epsilon_+} ^{ (1)}\left(\boldsymbol{r},\boldsymbol{r}\right) \equiv \langle \boldsymbol{r} | \delta \hat G_{\epsilon_+} ^{ (1)} | \boldsymbol{r} \rangle$ is given by
\begin{equation}\label{eq:Green function to first order definition}
 \delta G_{\epsilon_+} ^{ (1)}\left(\boldsymbol{r},\boldsymbol{r}\right)=\int d\boldsymbol{r}_{1}d\boldsymbol{r}_{2} G_{0,{\epsilon_+}}\left(\boldsymbol{r},\boldsymbol{r}_{1}\right)V_{12} \,  
 G_{0,{\epsilon_+}}\left(\boldsymbol{r}_{2},\boldsymbol{r}\right)   
\end{equation}
with $V_{12}
\equiv \langle \boldsymbol{r}_{2} |{\hat V} | \boldsymbol{r}_{1} \rangle$. Every term in that expression is a $4\times 4$ matrix. The element $\langle K'  \delta \hat{G}_{\epsilon_+}^{(1)}  K \rangle$, along with the trace from (\ref{eq:greens_function_all_valley_contribution2}), results in the first-order correction $\delta\rho_{\Delta K}^{(1)} (\boldsymbol{r})$ of the local electronic density (\ref{eq:local density and local density of states}).

The next subsections present two instances for an adatom and a vacancy, stressing potential properties tied to zero-mode topological characteristics.

\subsubsection{Valley coupling electronic density: adatom}

Here, we apply the previously described method to compute the first-order term $\delta \rho_{\text{ad}}^{\left(1\right)}\left(\boldsymbol{r}\right)$ resulting from valley coupling due to an adatom \cite{Dutreix2019}, highlighting essential computational steps and the outcome (further details in Appendix A).
 
The adatom potential ${\hat V}_{\text{ad}}$ in (\ref{eq:adatom_low_energy_sublattice}) is given by (\ref{Htopo}) with just one non-zero diagonal element $H_{aa}$. Furthermore, the two valleys in ${H}_0$ from (\ref{eq:valley_basis_Hamiltonian}) remain uncoupled, meaning the Green's function does not involve valley coupling terms (refer to Table \ref{table:sublattice and valley basis}). 
Then, to determine valley coupling contributions to  $\delta \rho_{\text{ad}}^{\left(1\right)}\left(\boldsymbol{r}\right)$, we need to compute the product $\delta \hat G_{\epsilon_+} ^{ (1)} \equiv \hat{G}_{0,\epsilon_+}{\hat V}_{\text{ad}}\hat{G}_{0,\epsilon_+}$. Given the expression of the adatom potential and evaluating the trace operation in (\ref{eq:greens_function_all_valley_contribution2}) over sublattices, the two valley coupling terms are 
\begin{align}\label{eq:adatom_scattring_first_order_valley}
     \delta \hat{G}^{KK'}_{AA} &=\hat{G}^{K}_{0,AA}\hat{V}^{KK'}_{AA}\hat{G}^{K'}_{0,AA}\\
      \delta\hat{G}^{KK'}_{BB} &=\hat{G}^{K}_{0,BA}\hat{V}^{KK'}_{AA}\hat{G}^{K'}_{0,AB}\nonumber
 \end{align}
as calculated in Appendix A. The energy integral from (\ref{eq:local density and local density of states}) is evaluated up to the Fermi energy $\epsilon_F$, adjusted by a gate bias voltage $V_g$, yielding $\epsilon_F = v_F |k_F|$ where $k_F \propto \sqrt{|V_g|}$. The voltage $V_g$ is sufficiently large to disrupt particle-hole symmetry by the potential $V_0$. Finally, 
\begin{equation}   
 \label{interfad}
\delta \rho_{\text{ad}}^{\left(1\right)}\left(\boldsymbol{r}\right) =   f_A \left(k_F r\right) \, \cos\chi_0 (\boldsymbol{r}) 
 -f_B \left(k_F r\right) \,\cos\chi_2 (\boldsymbol{r}) 
\end{equation}
where $ \chi_n (\boldsymbol{r}) \equiv \Delta\boldsymbol{K}\cdot\boldsymbol{r}+ n \, \theta$ and, 
\begin{align}
f_{A,B}\left(k_{F}  r\right) & \equiv \frac{a^2 \, V_0}{8 \pi v_F r^3}\int_{0}^{k_F r}dz\, z^{2} \, \text{Im}\, \left[H_{0,1}^{(1)}(z)\right]^2 \, . 
\end{align}
Figure \ref{fig:fig6} shows that the local electronic density change $\delta \rho_{\text{ad}}^{\left(1\right)}\left(\boldsymbol{r}\right)$ reveals $2 k_F  r$ Friedel oscillations \cite{Dutreix2019}. This oscillation typically comes with decay over long distances, expressed as $\delta\rho(\boldsymbol{r})\propto r^{-\alpha} \cos(2k_F \, r)$, where $\alpha$ is related to spatial dimensionality, an expected response of a fermion gas to a local perturbation. Valley coupling changes this anticipated oscillatory behavior by adding $\chi_n (\boldsymbol{r})$ oscillations due to the local adatom disturbance.
 A local potential creating topological states significantly alters Friedel oscillations due to protected chiral symmetry, as addressed in the next subsection \cite{Mariani2007, Cheianov2006, Abulafia2023}. 
 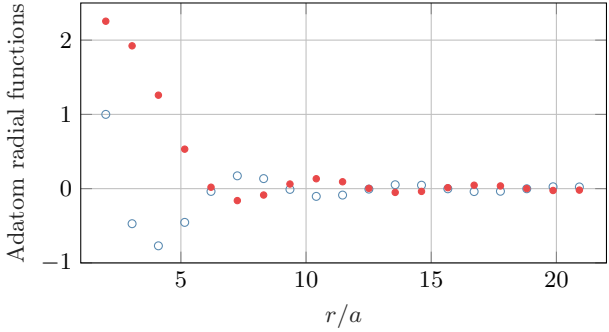
\begin{figure}
    \centering
\setlength\fheight{0.4\columnwidth}
\setlength\fwidth{0.85\columnwidth}
\input{pics/figure7/fradatom.tikz}
    \caption{Numerical evaluation of $f_{A}(r)/f_{0}$ (blue, empty), $f_{B}(r)/f_{0}$ (red, full)  normalized by $f_{0} \equiv f_{A}(r_0/a)$, $r_0/a=2$ with  $v_F=1$, so that $\epsilon_F = v_F\:k_F=0.5/a $ have units of $1/a$. Friedel oscillations of wavelength $\lambda_F=6.28a$ are observed. 
    }
    \label{fig:fig6}
\end{figure}
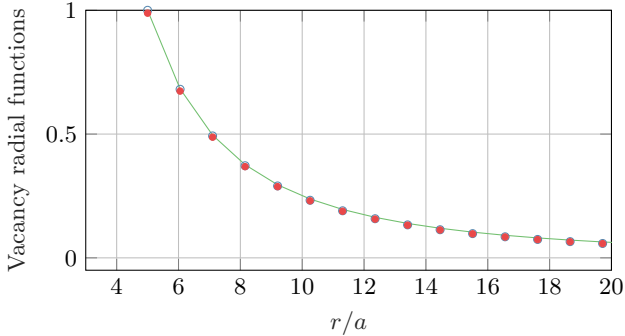
\begin{figure}
    \centering
\setlength\fheight{0.4\columnwidth}
\setlength\fwidth{0.85\columnwidth}
\input{pics/figure7/frvacancy.tikz}
    \caption{The spatial decay of $I_{A}(r)/I_{0}$ (blue, empty) and $I_{B}(r)/I_{0}$ (red, full) is numerically evaluated, where $I_{0}$ is normalized as $I_{0}=I_{A}(r_0/a)$ with $r_0/a=5$. This simplifies fitting the radial decay $(r_0/r)^2$ (green), with $v_F=1$.
    }
    \label{fig:fig7}
\end{figure}
%================ Ending Figure 1===========
 
\subsubsection{Valley coupling electronic density: vacancy}\label{subsubsec:vacancy_rho}

We use the previously mentioned method to calculate the first-order electronic density change $\delta \rho_{\text{vac}}^{(1)}\left(\boldsymbol{r}\right)$ due to a vacancy. This is compared to the local density correction (\ref{interfad}) for an adatom potential to differentiate between topological vacancies and non-topological adatoms.
 
 The vacancy potential ${\hat V}_{\text{vac}}$ in (\ref{calV}) is of the form (\ref{Htopo}) with inter-sublattice terms only.
To identify contributions from valley coupling to $\delta \rho_{\text{vac}}^{\left(1\right)}\left(\boldsymbol{r}\right)$, we need to calculate the product $\delta \hat G_{\epsilon_+} ^{ (1)} \equiv \hat{G}_{0,\epsilon_+}\hat{V}_{\text{vac}}\hat{G}_{0,\epsilon_+}$. Using the vacancy potential expression and the trace over sublattices in (\ref{eq:greens_function_all_valley_contribution2}), the valley-coupling contributions are found as,
\begin{align}\label{eq:first order valley and sublattice contributions}
\delta \hat{G}_{AA}^{KK'}  & =\hat{G}_{0,AA}^{K}\hat{V}_{AB}^{KK'}\hat{G}_{0,BA}^{K'}+\hat{G}_{0,AB}^{K}\hat{V}_{BA}^{KK'}\hat{G}_{0,AA}^{K'}\nonumber\\
\delta\hat{G}_{BB}^{KK'}  & =\hat{G}_{0,BB}^{K}\hat{V}_{BA}^{KK'}\hat{G}_{0,AB}^{K'}+\hat{G}_{0,BA}^{K}\hat{V}_{AB}^{KK'}\hat{G}_{0,BB}^{K'} \,,
\end{align}
where we dropped $(1)$ to shorten the notation.
These terms are calculated in Appendix B. To first order, the variation in local electronic density $\delta\rho_\text{vac}^{(1)}\left(\boldsymbol{r}\right)$ is derived from equation (\ref{eq:local density and local density of states}). This involves considering the imaginary part by restoring the imaginary energy $(\epsilon \rightarrow \epsilon_+ )$ and performing integration over energies. The upper limit for integration is set by the Fermi energy $\epsilon_F=0$, reflecting particle-hole and chiral symmetries, while the lower limit is determined by the bandwidth $\epsilon_{\text{min}}=-3t$. This results in contributions from two sublattices,
\begin{align}\label{eq:IAIB}
     I_{A}(r)=&\frac{a^2}{4\pi r^4}\int_{-2r/a}^{k_Fr}dz \,  \text{Im} \,H_{0}^{\left(1\right)}\left(z\right)\,z^2\frac{d}{dz}\left(z\frac{d}{dz}H_{0}^{\left(1\right)}\left(z\right)\right)\nonumber\\
     I_{B}(r)=&\frac{a^2}{4\pi r^4}\int_{-2r/a}^{k_Fr} dz \:z^3 \, \text{Im} \left(\frac{d}{dz}H_{0}^{\left(1\right)}\left(z\right)\right)^{2} \, .
\end{align}
Fig.\ref{fig:fig7} shows that integrals $I_{A}(r)$ and $I_{B}(r)$ are equal $I_{A} = I_{B} \equiv I(r)$, decay as $1/r^2$ near the vacancy and more quickly at greater distances. The change to first order of local electronic density associated with a vacancy is thus given by,
\begin{align}\label{eq:local density vacancy}
\delta\rho_{\text{vac}}^{(1)}\left(\boldsymbol{r}\right) =I(r)\left[\cos\left(\Delta\boldsymbol{K}\cdot\boldsymbol{r}\right)-\cos\left(\Delta\boldsymbol{K}\cdot\boldsymbol{r}+2\theta\right)\right] 
\, .
\end{align}
Unlike the adatom potential where the change of local electronic density (\ref{interfad}) involves two distinct radial amplitudes $f_{A,B}$, the vacancy potential involves only $I(r)$ as a result of particle-hole and chiral symmetries. Notably, the result from (\ref{eq:local density vacancy}) remains unaffected by the Fermi energy choice, whether set to zero or any non-zero value due to gating. A key point is that a vacancy preserves particle-hole symmetry irrespective of the Fermi energy selection. However, $I(r)$ is influenced by both the Fermi energy and the lower boundary selection.
By preserving both particle-hole and chiral symmetries, the vacancy potential eliminates Friedel oscillations; however, these oscillations appear when an adatom is present and breaks the particle-hole symmetry.    

The expression (\ref{eq:local density vacancy}) for $\delta\rho_{\text{vac}}^{(1)}\left(\boldsymbol{r}\right)$ has been derived to first order. It can be extended to incorporate all higher-order terms within the series (\ref{eq:local density and local density of states}), ultimately yielding the same result with a modified radial decay (See Appendix C).

 \section{Discussion}

This paper highlights the significance of anti-unitary symmetries and chiral symmetry in configuring defect potentials to satisfy topological classification requirements. We analyzed the structure of Hamiltonians to methodically demonstrate how to identify chiral symmetry and find when its off-diagonal components give rise to a non-zero analytical index.

By utilizing adatoms and vacancies as case studies, we presented a method for deriving defect potentials that maintain chiral symmetry, an essential requirement for topological classification. Using a relation between topological zero modes and the analytical index, we  calculated their number. Our method can be applied to topological textures, which involve non-local modifications of bond strengths specifically designed to produce a finite topological index. One prominent example of such a texture is the Kekulé distortion \cite{Hou2007,Jackiw2007}, known for generating topological zero modes and charge fractionalization. Interestingly, this texture shares the same potential matrix structure as a vacancy in terms of its non-vanishing elements, further highlighting the central role of the Hamiltonian's matrix structure in determining topological properties.

An extension to this involves examining several defects. When discussing two vacancies, each alters the Hamiltonian's off-diagonal components by adding extra matrix elements to the potential, thus increasing the number of zero modes localized on the opposite sublattice. The overall analytical index depends on the configuration, as type-$A$ vacancies and type-$B$ vacancies have opposite contributions. An $A+B$ setup gives a zero index, whereas $A+A$ or $B+B$ setups have an index of $\pm 2$. 

In a situation that includes both a vacancy and an adatom, the vacancy modifies the off-diagonal terms, whereas the adatom provides a diagonal mass term, disrupting chiral symmetry. Even though chiral symmetry is absent, the analytical index remains well defined and is influenced only by the vacancy, resulting in the number of topological zero modes staying the same as in the scenario with just a single vacancy.

We determined the electronic charge density using the Green's function, uncovering distinctive behaviors for each defect type. Although we did not consider other physical observables in our analysis, transport properties, for instance, can be directly obtained from the Green's function, suggesting a natural extension of this work.

Additional physical consequences of underlying topological features remain unexplored beyond the recognized role of edge states. The unresolved issues of whether Friedel oscillations manifest away from topology-inducing localized defects and the mechanisms of magnetic moment formation require further investigation. It must be emphasized that initial studies on graphene with vacancies predominantly addressed magnetic characteristics, with significant theoretical and experimental efforts dedicated to assessing its magnetic moments \cite{Yazyev2007,Ugeda2010,Lehtinen2004,Yuchen2004,Vozmediano2005,Yazyev2008,Palacios2008,Yazyev2010,Jiang2018,González-Herrero2016,Nair2013,Faccio2010,Bhatt2022,Valencia2017,Yndurain2014,Nair2012}, all within the context of the repulsive Hubbard model \cite{Hubbard1963,Vozmediano2005,Palacios2008}. Nonetheless, the topological nature of these defects had yet to be identified and therefore was not the focus of prior research.
\begin{acknowledgments}
 This research was funded by the Israel Science Foundation Grant No.~772/21 and the Pazy Foundation. \\
\end{acknowledgments}

%--------------------------------------------------------
\appendix
\section{Valley coupling electronic density: adatom}

In this appendix, further details are given about the calculation of the valley coupling change (\ref{interfad}) of electronic density $\delta \rho_{\text{ad}}^{\left(1\right)}$ induced by an adatom.  

We start from (\ref{eq:adatom_scattring_first_order_valley}) and use the notation, 
\begin{equation}
 \delta \hat{G}^{KK'}_{AA} \equiv \langle AK' | \delta \hat G_{\epsilon_+} | AK \rangle   
\end{equation} 
while omitting $\left(1\right)$ that stands for first-order and complex valued energies $\epsilon_+$ to shorten  notations. Using (\ref{eq:graphene Green function valley_K}), 
 \begin{equation} 
 {G}^{K}_{0,AA} \left(\boldsymbol{r}\right)\equiv \langle A |\hat G_{0,\epsilon}^{K }\left(\boldsymbol{r}\right) | A\rangle = e^{i\boldsymbol{K }\cdot\boldsymbol{r}} \frac{\epsilon}{v_{F}}{g} (\epsilon  r) \, . 
 \end{equation}
Positioning the adatom at the origin, $\boldsymbol{R_0} = \boldsymbol{0}$, along with,
\begin{equation}
  \langle \boldsymbol{r'} |\hat{V}^{KK'}_{AA} | \boldsymbol{r} \rangle = a^2V_0 \,\delta(\boldsymbol{r})\, \delta (\boldsymbol{r} - \boldsymbol{r'} ) \, ,   
\end{equation} leads to the valley coupling term, 
\begin{align}
\text{Tr}_{AB}  \delta G_{\epsilon_+}^{KK'}\left(\boldsymbol{r},\boldsymbol{r}\right) &= a^2V_0 
    {G}^{K}_{0,AA}\left(\boldsymbol{r} \right)  {G}^{K'}_{0,AA}\left( -\boldsymbol{r}\right)
    \\ \nonumber 
      &+a^2 V_0 {G}^{K}_{0,BA}\left(\boldsymbol{r}\right)  {G}^{K'}_{0,AB}\left(-\boldsymbol{r}\right) 
\end{align}
after adding up the two terms from (\ref{eq:adatom_scattring_first_order_valley}) while taking the trace over sublattices, and then substituting the unperturbed Green's function. Considering that ${g} (\epsilon  r)$ is symmetric with respect to $r$,
\begin{equation}
   \hat L^{\dagger}{g} (-\epsilon \, r)=-\hat L^{\dagger}{g} (\epsilon \, r) \, , 
\end{equation}
leads to
\begin{align}
\text{Tr}_{AB}  \delta G_{\epsilon}^{KK'} 
= a^2V_0 \, 
    e^{i\Delta\boldsymbol{K}\cdot\boldsymbol{r}} 
      \left[\frac{\epsilon^2}{v_{F}^2}{g}^2 (\epsilon \, r)+(\hat L^\dagger{g} (\epsilon \, r) )^2\right] \, . 
\end{align}
Using (\ref{Lang}), this expression splits into two terms, each one with a different angular behavior, 
\begin{equation}
   \text{Tr}_{AB}\delta G_{\epsilon} ^{KK'}\left(\boldsymbol{r},\boldsymbol{r}\right) \equiv e^{i\Delta\boldsymbol{K}\cdot\boldsymbol{r}}\left[h_{A}\left(\epsilon,r\right)-e^{2i\theta} h_{B}\left(\epsilon,r\right)\right] 
\end{equation}
where,
\begin{equation}
   \begin{cases}
h_{A}\left(\epsilon,r\right)=a^{2}V_{0}\frac{\epsilon^{2}}{v_{F}^{2}}g^{2}(\epsilon\,r)\\
h_{B}\left(\epsilon,r\right)=a^{2}V_{0}(\frac{d}{dr}g(\epsilon\,r))^{2}.
\end{cases} 
\end{equation}
We similarly determine the opposite valley scattering and combine both results to obtain (\ref{interfad}) announced in the text.

\section{Valley coupling electronic density: vacancy}

In this appendix, further details are given about the calculation of the valley coupling change (\ref{eq:local density vacancy}) of electronic density $\delta \rho_{\text{vac}}^{\left(1\right)}$ induced by a vacancy.  

We start from (\ref{eq:first order valley and sublattice contributions}). To evaluate it, we note that 
each term in (\ref{eq:Green function to first order definition}) involves spatially dependent matrix elements of the vacancy potential.  
Denoting 
\begin{equation}
    V_{12} 
=   \langle \boldsymbol{r}_{1} | { \hat{V}_{\text{vac}}} | \boldsymbol{r}_{2} \rangle \equiv a^2  v_F A_{12} \, \delta\left(\boldsymbol{r}_{1}-\boldsymbol{r}_{2}\right)
\end{equation} then,
\begin{align} 
A_{12}& =\left(
\begin{smallmatrix}
0 & 0 & 0 & \delta\left(\boldsymbol{r}_{1}\right)\hat L_{{2}}^{\dagger}\\
0 & 0 & -\delta\left(\boldsymbol{r}_{1}\right)\hat L_{{2}} & 0\\
0 & \delta\left(\boldsymbol{r}_{2}\right)\hat L_{{1}}^{\dagger} & 0 & 0\\
-\delta\left(\boldsymbol{r}_{2}\right) \hat L_{{1}} & 0 & 0 & 0
\end{smallmatrix}\right) 
\end{align}
where $\hat L_i$ stands for $\hat L$ defined in (\ref{Lop}) and evaluated at $\boldsymbol{r}_{i}$. Upon substituting into (\ref{eq:first order valley and sublattice contributions}), integrating by parts to handle derivatives of $\delta\left(\boldsymbol{r}_1-\boldsymbol{r}_2\right)$, and using 
\begin{equation}
   \hat L_{{2}}^{\dagger} \, G_{0,BA}^{K'}\left(\boldsymbol{r}_{2}-\boldsymbol{r}_1\right) = -\hat L_{1}^{\dagger} \, G_{0,BA}^{K'}\left(\boldsymbol{r}_{2}-\boldsymbol{r}_1\right) 
\end{equation} 
leads to 
\begin{align}
\delta G_{AA}^{KK'}\left(\boldsymbol{r},\boldsymbol{r}'\right) & =a^2v_{F} \; \{ G_{0,AA}^{K}\left(\boldsymbol{r}\right) \hat L'^{\dagger}\, G_{0,BA}^{K'}\left(-\boldsymbol{r}'\right)\nonumber\\& +\hat L^{\dagger} \, G_{0,AB}^{K}\left(\boldsymbol{r}\right)\;G_{0,AA}^{K'}\left(-\boldsymbol{r}'\right) \} \nonumber\\ 
\delta G_{BB}^{KK'}\left(\boldsymbol{r},\boldsymbol{r}'\right)  & =a^2v_{F}\; \{ \hat L^{\dagger} \, G_{0,BB}^{K}\left(\boldsymbol{r}\right)\;G_{0,AB}^{K'}\left(-\boldsymbol{r}'\right)\nonumber\\ &+ \;G_{0,BA}^{K}\left(\boldsymbol{r}\right)\;\hat L'^{\dagger} \, G_{0,BB}^{K'}\left(-\boldsymbol{r}'\right) \} 
\end{align}
where $\hat L' \equiv -i\partial_{x'}-\partial_{y'}$ 
is evaluated at $\boldsymbol{r'}$. Since ${g} (\epsilon \, r)$ in  (\ref{eq:graphene Green function valley_K}) 
is even, then $\hat L^{\dagger}{g} (-\epsilon \, r)=-\hat L^{\dagger}{g} (\epsilon \, r)$, leading to further simplifications.
Summing over sublattices $A$ and $B$ in (\ref{eq:greens_function_all_valley_contribution2}), 
the resulting Green's function variation to first order is 
\begin{align}
\text{Tr}_{AB}\delta G_{\epsilon} ^{KK'} =&-2 a^2  e^{i\Delta\boldsymbol{K}\cdot\boldsymbol{r}} 
\epsilon \left[{g}\left(\epsilon r\right)\nabla^{2} {g}\left(\epsilon r\right)+\left(\hat{L}^{\dagger} \, {g} \left(\epsilon  r\right)\right)^{2}\right] \, .
\end{align}
Relations (\ref{Lang}) and $\hat L^{\dagger}\, \hat L^{\vphantom{\ast}}= \hat L^{\vphantom{\ast}}\, \hat L^{\dagger}=-\nabla^{2}$,  
provide two angular contributions, one from each sublattice, so that, 
\begin{equation}\label{eq:green_KK'}
\text{Tr}_{AB} \, \delta G_{\epsilon} ^{KK'} = \, e^{i\Delta\boldsymbol{K}\cdot\boldsymbol{r}}\left[\tilde f_A\left(\epsilon,r\right)-\tilde f_B\left(\epsilon,r\right)\: e^{2 i\theta}\right] \, .
\end{equation}
Using (\ref{K'K}) allows to obtain the second intervalley contribution $(K \leftrightarrow K' )$,
\begin{equation}\label{eq:green_K'K}
\text{Tr}_{AB} \, \delta G_{\epsilon} ^{K'K} = \, e^{-i\Delta\boldsymbol{K}\cdot\boldsymbol{r}}\left[\tilde f_A\left(\epsilon,r\right)-\tilde f_B\left(\epsilon,r\right)\: e^{-2 i\theta}\right]
\end{equation}
with 
\begin{equation}
\begin{cases}
\tilde f_A(\epsilon_,r) \equiv -2 a^2\epsilon \, {g} (\epsilon \, r)\frac{1}{r}\frac{d}{dr}\left(r\frac{d}{dr}{g} (\epsilon \, r)\right) \\
\tilde f_B\left(\epsilon,r\right) \equiv -2 a^2\epsilon \, \left(\frac{d}{dr}\, {g} (\epsilon \, r)\right)^{2},
\end{cases}
\end{equation}
and the two integrals appearing in (\ref{eq:IAIB}) are calculated accordingly, 
\begin{equation}
 I_{A,B}(r)= -\frac{2}{\pi} \int_{-3t}^{\epsilon_F}d\epsilon \, \text{Im} \tilde f_{A,B}\left(\epsilon,r\right).
\end{equation}
hence leading to (\ref{eq:local density vacancy}).

\begin{figure}
    \centering
 \hspace{-10pt}
\setlength\fheight{0.4\columnwidth}
\setlength\fwidth{0.85\columnwidth}
\input{pics/figure7/vacancyoscillations.tikz}
    \caption{Numerical evaluation of  $I_{A}(r)/I_{0}$ (blue, empty), $I_{B}(r)/I_{0}$ (red, full) spatial decay using the normalization $I_{0}=I_{A}(r_0/a)$, $r_0/a=2.825$ to simplify the radial decay fit $(r_0/r)^2$ (green) and $v_F=1$. 
The oscillations have a wavelength smaller than the lattice constant $a$ and thus were not considered in the main text, where we assumed $k r \gg 1$. The observed $1/r^2$ decay and the subatomic oscillations match the results in \cite{Ugeda2010}.
    }
    \label{fig:fig8}
\end{figure}
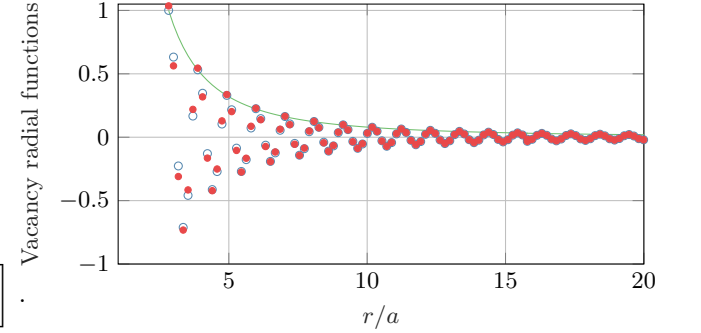

\section{\texorpdfstring{Higher order corrections to $\delta\rho_{\text{vac}}$ }{Higher order corrections to vac }}

In this appendix, we derive expression (\ref{eq:local density vacancy}) for the modification of the local electronic density, going beyond first-order perturbation.  It is shown that each higher order contribution is proportional to $\delta\rho_{\text{vac}}^{(1)}\left(\boldsymbol{r}\right)$ up to a radial function. We consider only odd order terms since they contribute to inter-valley scattering, a property of the potential ${\hat V}_{\text{vac}}$  which contains only $K\to K'$ coupling terms. For graphene with a vacancy,  Green's functions $\hat{G}_{0,AB/BA}^{K/K'}$ vanish when inserted between two local $\hat{V}_{\text{vac}}$ operators,
\begin{equation}
\langle\boldsymbol{r}|\hat{G}_{0,ij}^{\alpha}\hat{V}_{jk}^{\alpha\beta}\hat{G}_{0,kl}^{\beta }\hat{V}_{lm}^{\beta\gamma  }\hat{G}_{0,mn}^{\gamma }|\boldsymbol{r}'\rangle\propto\hat{G}_{0,kl}^{\beta }\left(0\right)
\end{equation}
where we used Greek letters $\alpha,\beta,\gamma$ to denote valley indexes $K,K'$, and Latin letters $i,j,k,m,n$ to denote sublattice $A,B$. For $k\neq l$, this expression vanishes since  $G_{0,AB/BA}^{\beta}(0)=0$. Thus, all scattering processes can be divided into contributions from the first-order term multiplied by a recurring sequence of operators. For instance, the typical $AA$ scattering term is
\begin{widetext}
\begin{equation}\label{eq:Green function block form}
\delta \hat{G}_{AA}^{KK'\left(2n+1\right)} =\left(\hat{G}_{0,AA}^{K}\hat{V}_{AB}^{KK'}\hat{G}_{0,BB}^{K'}\hat{V}_{BA}^{K'K}\right)^{n}\hat{G}_{0,AA}^{K}\hat{V}_{AB}^{KK'}\hat{G}_{0,BA}^{K'}+\hat{G}_{0,AB}^{K}\hat{V}_{BA}^{KK'}\hat{G}_{0,AA}^{K'}\left(\hat{V}_{AB}^{K'K}\hat{G}_{0,BB}^{K}\hat{V}_{BA}^{KK'}\hat{G}_{0,AA}^{K'}\right)^{n}
\end{equation}
\end{widetext}
The repeating terms, e.g.  $\hat{G}_{0,AA}^{K}\hat{V}_{AB}^{KK'}\hat{G}_{0,BB}^{K'}\hat{V}_{BA}^{K'K}$, are always composed of products of four terms: two diagonal elements of the sublattice $\hat{G}_{0,AA/BB}$ which do not contain $\hat{L}$ or $\hat{L}^{\dagger}$ (\ref{eq:graphene Green function valley_K}) and another two Hermitian conjugate operators, such that for each $\hat{L}$ operator there is an $\hat{L}^{\dagger}$ operator. These pairs of $\hat{L}$, $\hat{L}^{\dagger}$ operators cancel the $\theta$ dependence (\ref{Lang}) of the repeating blocks. Finally, we obtain the matrix elements,
\begin{widetext}
\begin{align}\label{eq:block_first_order}
\left\langle \boldsymbol{r}\right|\hat{G}_{0,AA}^{K}\hat{V}_{AB}^{KK'}\hat{G}_{0,BB}^{K'}\hat{V}_{BA}^{K'K}\left|\boldsymbol{r}'\right\rangle = & \left(a^{2}v_{F}\right)^{2}G_{0,AA}^{K}\left(\boldsymbol{r}\right)\left[\hat{L}'^{\dagger}\hat{L}'G_{0,BB}^{K'}\left(\boldsymbol{r}'\right)\right]\delta\left(\boldsymbol{r}'\right)\\
\left\langle \boldsymbol{r}\right|\hat{V}_{AB}^{K'K}\hat{G}_{0,BB}^{K}\hat{V}_{BA}^{KK'}\hat{G}_{0,AA}^{K'}\left|\boldsymbol{r}'\right\rangle = & \left(a^{2}v_{F}\right)^{2}\delta\left(\boldsymbol{r}\right)\left[\hat{L}^{\dagger}\hat{L}G_{0,BB}^{K}\left(\boldsymbol{r}\right)\right]G_{0,AA}^{K'}\left(\boldsymbol{r}'\right)\nonumber 
\end{align}
and their relation to the $(2n+1)^{\text th}$ term 
\begin{align}\label{'eq:higer order blocks'}
\left\langle \boldsymbol{r}\right|\left(\hat{G}_{0,AA}^{K}\hat{V}_{AB}^{KK'}\hat{G}_{0,BB}^{K'}\hat{V}_{BA}^{K'K}\right)^{n}\left|\boldsymbol{r}'\right\rangle  & =\left(C^{K'}\right)^{n-1}\left\langle \boldsymbol{r}\right|\hat{G}_{0,AA}^{K}\hat{V}_{AB}^{KK'}\hat{G}_{0,BB}^{K'}\hat{V}_{BA}^{K'K}\left|\boldsymbol{r}'\right\rangle \\
\left\langle \boldsymbol{r}\right|\left(\hat{V}_{AB}^{K'K}\hat{G}_{0,BB}^{K}\hat{V}_{BA}^{KK'}\hat{G}_{0,AA}^{K'}\right)^{n}\left|\boldsymbol{r}'\right\rangle  & =\left(C^{K}\right)^{n-1}\left\langle \boldsymbol{r}\right|\hat{V}_{AB}^{K'K}\hat{G}_{0,BB}^{K}\hat{V}_{BA}^{KK'}\hat{G}_{0,AA}^{K'}\left|\boldsymbol{r}'\right\rangle \nonumber 
\end{align}
where we have defined two dimensionless scaling functions, 
\begin{equation}
    C^{K'/K}\left(\epsilon \right
)\equiv\left(a^{2}v_{F}\right)^2\left.\left[\hat{L}_{{1}}^{\dagger}\hat{L}_{{1}}^{\vphantom{\ast}}G_{0,BB}^{K'/K}\left(\boldsymbol{r}_{1}\right)\right]\right|_{\boldsymbol{r}_{1}=0}G_{0,AA}^{K/K'}\left(0\right).
\end{equation}
 These two functions are equal for   $\left|\boldsymbol{K}\right|=\left|\boldsymbol{K}'\right|$, since the first derivative  $\hat{L}_{1}G_{0,BB}^{K/K'}\left(\boldsymbol{r}_1\right)$ vanishes at the origin. We examine the $(2n+1)$-contribution to the Green's function. We use \eqref{'eq:higer order blocks'} to calculate the matrix elements of \eqref{eq:Green function block form}
     \begin{align}
\left\langle \boldsymbol{r}\right|\delta \hat{G}_{AA}^{KK'\left(2n+1\right)}\left|\boldsymbol{r}'\right\rangle  & =\left(C^{K}\right)^{n-1}\int d\boldsymbol{r}_{1}\left\langle \boldsymbol{r}\right|\hat{G}_{0,AA}^{K}\hat{V}_{AB}^{KK'}\hat{G}_{0,BB}^{K'}\hat{V}_{BA}^{K'K}\left|\boldsymbol{r}_{1}\right\rangle \left\langle \boldsymbol{r}_{1}\right|\hat{G}_{0,AA}^{K}\hat{V}_{AB}^{KK'}\hat{G}_{0,BA}^{K'}\left|\boldsymbol{r}'\right\rangle \\
 & +\left(C^{K}\right)^{n-1}\int d\boldsymbol{r}_{1}\left\langle \boldsymbol{r}\right|\hat{G}_{0,AB}^{K}\hat{V}_{BA}^{KK'}\hat{G}_{0,AA}^{K'}\left|\boldsymbol{r}_{1}\right\rangle \left\langle \boldsymbol{r}_{1}\right|\hat{V}_{AB}^{K'K}\hat{G}_{0,BB}^{K}\hat{V}_{BA}^{KK'}\hat{G}_{0,AA}^{K'}\left|\boldsymbol{r}'\right\rangle \nonumber \, .
\end{align}
Integrating over $\boldsymbol{r}_1$ 
and inserting all the matrix elements previously calculated provides an identical contribution as the first order multiplied by the  $\left(C^{K}\right)^{n}$ coefficient,
\begin{align}
\delta G_{AA}^{KK'\left(2n+1\right)}\left(\boldsymbol{r},\boldsymbol{r}'\right) & =\left(C^{K}\right)^{n}a^{2}\epsilon \,  e^{i\boldsymbol{K}\cdot\boldsymbol{r}}{g} (\epsilon \, r)\left(\hat{L}'^{\dagger}e^{-i\boldsymbol{K}'\cdot\boldsymbol{r}'}\hat{L}'^{\vphantom{\ast}}{g} (-\epsilon \, r')\right)\\
 & +\left(C^{K}\right)^{n}a^{2}\epsilon \, e^{-i\boldsymbol{K}'\cdot\boldsymbol{r}'}\left(\hat{L}^{\dagger}e^{i\boldsymbol{K}\cdot\boldsymbol{r}}\hat{L}^{\vphantom{\ast}}{g} (\epsilon \, r)\right){g} (-\epsilon \, r') \nonumber \, .
\end{align}
An analogous calculation for sublattice $B$ provides, 
\begin{align}
\delta {G}_{BB}^{KK'\left(2n+1\right)} \left(\boldsymbol{r},\boldsymbol{r}'\right) & =\left(C^{K}\right)^{n}a^{2}\epsilon \;e^{i\boldsymbol{K}\cdot\boldsymbol{r}}\left(\hat{L}^{\dagger}g\left(\epsilon\,r\right)\right)\left(\hat{L}'^{\dagger}e^{-i\boldsymbol{K}'\cdot\boldsymbol{r}'}g\left(-\epsilon\,r'\right)\right)\\
 & +\left(C^{K}\right)^{n}a^{2}\epsilon \,e^{-i\boldsymbol{K}'\cdot\boldsymbol{r}'}\left(\hat{L}^{\dagger}e^{i\boldsymbol{K}\cdot\boldsymbol{r}}g\left(\epsilon\,r\right)\right)\left(\hat{L}'^{\dagger}g\left(-\epsilon\,r'\right)\right).\nonumber
\end{align}
Taking $\boldsymbol{r}'=\boldsymbol{r}$ and summing over the two sublattices, we can split   into radial and angular behavior, 
\begin{equation}\label{eq:green_K'K_higher}
\text{Tr}_{AB} \, \delta G_{\epsilon} ^{K'K(2n+1)} = \, e^{-i\Delta\boldsymbol{K}\cdot\boldsymbol{r}}\left[ F_{A,2n+1}\left(\epsilon,r\right)- F_{B,2n+1}\left(\epsilon,r\right)\: e^{-2 i\theta}\right],
\end{equation}
where the radial functions of the $(2n+1)$ order are
\begin{equation}
\begin{cases}
 F_{A,2n+1}(\epsilon,r) \equiv -2 a^2\left(C^{K}\left(\epsilon\right)\right)^{n}\epsilon \, {g} (\epsilon \, r)\frac{1}{r}\frac{d}{dr}\left(r\frac{d}{dr}{g} (\epsilon \, r)\right) \\
 F_{B,2n+1}(\epsilon,r) \equiv -2 a^2\left(C^{K}\left(\epsilon\right)\right)^{n}\epsilon \, \left(\frac{d}{dr}\, {g} (\epsilon \, r)\right)^{2}.
\end{cases}
\end{equation}
 Integrating over energy provides two identical integrals to leading order in $a/r$, where we have used the asymptotic limit of the Hankel functions $\ensuremath{H_{\nu}^{(1)}\left(x\right)\simeq\sqrt{\frac{2}{\pi x}}e^{i(x-\nu\frac{\pi}{2}-\frac{\pi}{4})}}$ provided $u=\frac{a\epsilon}{v_F}$, such that the arguments in $H_{\nu}^{(1)}(u\frac{r}{a})$ is large throughout the integration range for $\epsilon_F=0$,
\begin{equation}
 I_{A,2n+1}(r)= -\frac{2}{\pi} \int_{-3t}^{\epsilon_F}d\epsilon \, \text{Im}  F_{A,2n+1}\left(\epsilon,r\right)\simeq\frac{1}{2\pi^2 r}\frac{1}{a}\int_{-1}^{k_Fa}du\left(C^{K}\left(u\right)\right)^{n}e^{2i(\frac{r}{a}u-\frac{\pi}{4})}u^{2}
\end{equation}
\begin{equation}
 \, I_{B,2n+1}(r)=- \frac{2}{\pi}\int_{-3t}^{\epsilon_F} d\epsilon \: \text{Im}  F_{B,2n+1}\left(\epsilon,r\right)\simeq \frac{1}{2\pi^2 r}\frac{1}{a}\int_{-1}^{k_Fa}du\left(C^{K}\left(u\right)\right)^{n}e^{2i(\frac{r}{a}u-\frac{\pi}{4})}u^{2}\, .
\end{equation}
 \end{widetext}
This leads to two equivalent radial functions for all orders of $n$, $I_{A,2n+1}(r)=I_{B,2n+1}(r)=I_{2n+1}(r)$, hence generalizing the first-order expression, namely 
\begin{align}
\delta\rho_{\text{vac}}\left(\boldsymbol{r}\right) & =F\left(r\right)\left[\cos\left(\Delta\boldsymbol{K}\cdot\boldsymbol{r}\right)-\cos\left(\Delta\boldsymbol{K}\cdot\boldsymbol{r}+2\theta\right)\right],
\end{align}
where $F\left(r\right)=\sum_{n=0}^{\infty}I_{2n+1}(r)$.

\bibliography{PRB.bib}% Produces the bibliography via BibTeX.

\end{document}

%% file: pics/figure7/fradatom.tikz
\definecolor{mycolor1}{rgb}{0.34667,0.53600,0.69067}%
\definecolor{mycolor2}{rgb}{0.91529,0.28157,0.28784}%
\definecolor{mycolor3}{rgb}{0.44157,0.74902,0.43216}%
\definecolor{mycolor4}{rgb}{1.00000,0.59843,0.20000}%
\begin{tikzpicture}

\begin{axis}[%
width=0.951\fwidth,
height=\fheight,
at={(0\fwidth,0\fheight)},
scale only axis,
xmin=1,
xmax=22,
xlabel style={font=\color{white!15!black}},
xlabel={$r/a$ },
ymin=-1,
ymax=2.5,
ylabel style={font=\color{white!15!black}},
ylabel={Adatom radial functions},
axis background/.style={fill=white},
xmajorgrids,
ymajorgrids,
ticklabel style={/pgf/number format/fixed}
]
\addplot [color=mycolor1, mark=o, only marks, mark options={fill=mycolor1},mark size=1.5, forget plot]
  table[row sep=crcr]{%
2.00000000000000	1\\
3.05100000000000	-0.471945594109866\\
4.10200000000000	-0.769729782630233\\
5.15300000000000	-0.454156770812953\\
6.20400000000000	-0.0372707944188011\\
7.25500000000000	0.172684720048970\\
8.30600000000000	0.135228205912637\\
9.35700000000000	-0.00997332370073318\\
10.4080000000000	-0.104071118755800\\
11.4590000000000	-0.0866008623309823\\
12.5100000000000	-0.00602581704551139\\
13.5610000000000	0.0525859482511119\\
14.6120000000000	0.0463444107375779\\
15.6630000000000	-0.00187442929952773\\
16.7140000000000	-0.0392863609685251\\
17.7650000000000	-0.0350385145612957\\
18.8160000000000	-0.00164261926638908\\
19.8670000000000	0.0252654196222616\\
20.9180000000000	0.0229217271035165\\
%21.9690000000000	-0.000928333247060988\\
%23.0200000000000	-0.0205977880042004\\
%24.0710000000000	-0.0186948006732101\\
%25.1220000000000	-0.000411995375582558\\
%26.1730000000000	0.0148800624918875\\	
%27.2240000000000	0.0135381733301206\\
%28.2750000000000	-0.000708496956403531\\
%29.3260000000000	-0.0127177835396598\\
};
\addplot [color=mycolor2, mark=,only marks, mark options={solid, fill=mycolor2, mycolor2}, mark size=1.2,forget plot]
  table[row sep=crcr]{%
2.00000000000000	2.25460184947871\\
3.05100000000000	1.92304068123535\\
4.10200000000000	1.25856933192210\\
5.15300000000000	0.531814145504109\\
6.20400000000000	0.0193050838386891\\
7.25500000000000	-0.160770621698089\\
8.30600000000000	-0.0858709548897174\\
9.35700000000000	0.0621418122283382\\
10.4080000000000	0.133467234089627\\
11.4590000000000	0.0926962628088011\\
12.5100000000000	0.00353027385121282\\
13.5610000000000	-0.0504861495976224\\
14.6120000000000	-0.0369437920018799\\
15.6630000000000	0.0130741240370001\\
16.7140000000000	0.0462452147165497\\
17.7650000000000	0.0365242532011371\\
18.8160000000000	0.000883265478812683\\
19.8670000000000	-0.0245194579126409\\
20.9180000000000	-0.0196378464007656\\
%21.9690000000000	0.00499748726006316\\
%23.0200000000000	0.0232134233028740\\
%24.0710000000000	0.0192447081681710\\
%25.1220000000000	8.83701323855194e-05\\
%26.1730000000000	-0.0145222989541067\\	
%27.2240000000000	-0.0120180105057667\\
%28.2750000000000	0.00261927654533229\\
%29.3260000000000	0.0139610515480563\\
};
\end{axis}

\begin{axis}[%
width=1.227\fwidth,
height=1.227\fheight,
at={(-0.16\fwidth,-0.135\fheight)},
scale only axis,
xmin=0,
xmax=1,
ymin=0,
ymax=1,
axis line style={draw=none},
ticks=none,
axis x line*=bottom,
axis y line*=left,
ticklabel style={/pgf/number format/fixed}
]
\end{axis}
\end{tikzpicture}%

%% file: pics/figure7/frvacancy.tikz
\definecolor{mycolor1}{rgb}{0.34667,0.53600,0.69067}%
\definecolor{mycolor2}{rgb}{0.91529,0.28157,0.28784}%
\definecolor{mycolor3}{rgb}{0.44157,0.74902,0.43216}%
\definecolor{mycolor4}{rgb}{1.00000,0.59843,0.20000}%
\begin{tikzpicture}

\begin{axis}[%
width=0.951\fwidth,
height=\fheight,
at={(0\fwidth,0\fheight)},
scale only axis,
xmin=3,
xmax=20,
xlabel style={font=\color{white!15!black}},
xlabel={$r/a$ },
ymin=-0.05,
ymax=1,
ylabel style={font=\color{white!15!black}},
ylabel={Vacancy radial functions},
axis background/.style={fill=white},
xmajorgrids,
ymajorgrids,
ticklabel style={/pgf/number format/fixed}
]
\addplot [color=mycolor1, mark=o, only marks, mark options={fill=mycolor1},mark size=1.5, forget plot]
  table[row sep=crcr]{%
5	1\\
6.05100000000000	0.680953897435997\\
7.10200000000000	0.492516865770316\\
8.15300000000000	0.372065353237620\\
9.20400000000000	0.290450730646880\\
10.2550000000000	0.232618992040103\\
11.3060000000000	0.190159634989650\\
12.3570000000000	0.158075130430817\\
13.4080000000000	0.133244712126329\\
14.4590000000000	0.113638177371645\\
15.5100000000000	0.0978886772632878\\
16.5610000000000	0.0850487033721438\\
17.6120000000000	0.0744447039514998\\
18.6630000000000	0.0655872708733218\\
19.7140000000000	0.0581138861634162\\
20.7650000000000	0.0517514063172606\\
21.8160000000000	0.0462908749980309\\
22.8670000000000	0.0415702428419482\\
23.9180000000000	0.0374622799470731\\
24.9690000000000	0.0338659715765050\\
26.0200000000000	0.0307002955696367\\
27.0710000000000	0.0278996568502717\\
28.1220000000000	0.0254104932991354\\
29.1730000000000	0.0231887217434566\\
};
\addplot [color=mycolor2, mark=,only marks, mark options={solid, fill=mycolor2, mycolor2}, mark size=1.2,forget plot]
  table[row sep=crcr]{%
5	0.988357179792706\\
6.05100000000000	0.673632191101653\\
7.10200000000000	0.487504027648348\\
8.15300000000000	0.368425050714866\\
9.20400000000000	0.287690737691943\\
10.2550000000000	0.230456582215883\\
11.3060000000000	0.188421124936846\\
12.3570000000000	0.156648046477354\\
13.4080000000000	0.132053059184699\\
14.4590000000000	0.112628782958827\\
15.5100000000000	0.0970232342735629\\
16.5610000000000	0.0842989209443851\\
17.6120000000000	0.0737892381864252\\
18.6630000000000	0.0650097179122812\\
19.7140000000000	0.0576014324022469\\
20.7650000000000	0.0512938979114452\\
21.8160000000000	0.0458801622916040\\
22.8670000000000	0.0411997094613503\\
23.9180000000000	0.0371264986297200\\
24.9690000000000	0.0335604486708991\\
26.0200000000000	0.0304212786649116\\
27.0710000000000	0.0276439880653931\\
28.1220000000000	0.0251754960746223\\
29.1730000000000	0.0229721125408027\\
};
\addplot [color=mycolor3, mark=, mark options={solid, fill=mycolor3, mycolor3}, forget plot]
  table[row sep=crcr]{%
5	1\\
6.05100000000000	0.682787721761600\\
7.10200000000000	0.495654065501359\\
8.15300000000000	0.376101550668198\\
9.20400000000000	0.295111945168295\\
10.2550000000000	0.237721618950787\\
11.3060000000000	0.195578921273664\\
12.3570000000000	0.163724591425118\\
13.4080000000000	0.139063131902871\\
14.4590000000000	0.119581361343013\\
15.5100000000000	0.103924133719599\\
16.5610000000000	0.0911521452870689\\
17.6120000000000	0.0805977012039724\\
18.6630000000000	0.0717756437845312\\
19.7140000000000	0.0643265862223424\\
20.7650000000000	0.0579797232152377\\
21.8160000000000	0.0525278661590643\\
22.8670000000000	0.0478103172098081\\
23.9180000000000	0.0437008903815186\\
24.9690000000000	0.0400993848175334\\
26.0200000000000	0.0369254183354494\\
27.0710000000000	0.0341139030555552\\
28.1220000000000	0.0316116817373716\\
29.1730000000000	0.0293749965756098\\
};
\end{axis}

\begin{axis}[%
width=1.227\fwidth,
height=1.227\fheight,
at={(-0.16\fwidth,-0.135\fheight)},
scale only axis,
xmin=0,
xmax=1,
ymin=0,
ymax=1,
axis line style={draw=none},
ticks=none,
axis x line*=bottom,
axis y line*=left,
ticklabel style={/pgf/number format/fixed}
]
\end{axis}
\end{tikzpicture}%

%% file: pics/figure7/vacancyoscillations.tikz
\definecolor{mycolor1}{rgb}{0.34667,0.53600,0.69067}%
\definecolor{mycolor2}{rgb}{0.91529,0.28157,0.28784}%
\definecolor{mycolor3}{rgb}{0.44157,0.74902,0.43216}%
\definecolor{mycolor4}{rgb}{1.00000,0.59843,0.20000}%
\begin{tikzpicture}

\begin{axis}[%
width=0.951\fwidth,
height=\fheight,
at={(0\fwidth,0\fheight)},
scale only axis,
xmin=1,
xmax=20,
xlabel style={font=\color{white!15!black}},
xlabel={$r/a$ },
ymin=-1.0,
ymax=1.05,
ylabel style={font=\color{white!15!black}},
ylabel={Vacancy radial functions},
axis background/.style={fill=white},
xmajorgrids,
ymajorgrids,
ticklabel style={/pgf/number format/fixed}
]
\addplot [color=mycolor1, mark=o, only marks, mark options={fill=mycolor1},mark size=1.5, forget plot]
  table[row sep=crcr]{%
2.82500000000000	1\\
3.00016666666667	0.632096044387791\\	
3.17533333333333	-0.226625320945366\\
3.35050000000000	-0.711118765712648\\
3.52566666666667	-0.458669047376308\\
3.70083333333333	0.166984175810623\\
3.87600000000000	0.531838009593714\\
4.05116666666667	0.346819078852980\\
4.22633333333333	-0.129386766111903\\
4.40150000000000	-0.412957244559394\\
4.57666666666667	-0.270634486351910\\
4.75183333333333	0.104065180009076\\
4.92700000000000	0.330077025670681\\
5.10216666666667	0.216492825201727\\
5.27733333333333	-0.0861271582939469\\
5.45250000000000	-0.269982151671500\\
5.62766666666667	-0.176688377182033\\
5.80283333333333	0.0729041895524451\\
5.97800000000000	0.225010709390488\\
6.15316666666667	0.146602026026073\\
6.32833333333333	-0.0628390913550676\\
6.50350000000000	-0.190471575016001\\
6.67866666666667	-0.123330229009109\\
6.85383333333333	0.0549731336678421\\
7.02900000000000	0.163362384438123\\
7.20416666666667	0.104975148776325\\
7.37933333333333	-0.0486888424252292\\
7.55450000000000	-0.141688790140610\\
7.72966666666667	-0.0902542618693793\\
7.90483333333333	0.0435737176518205\\
8.08000000000000	0.124084134002974\\
8.25516666666667	0.0782762514187778\\
8.43033333333333	-0.0393431101659531\\
8.60550000000000	-0.109586184927500\\
8.78066666666667	-0.0684061497146110\\
8.95583333333333	0.0357953698858342\\
9.13100000000000	0.0975016772613598\\
9.30616666666667	0.0601820225914409\\
9.48133333333333	-0.0327840755934566\\
9.65650000000000	-0.0873206949422689\\
9.83166666666667	-0.0532613017258768\\
10.0068333333333	0.0302008183590165\\
10.1820000000000	0.0786614965901357\\
10.3571666666667	0.0473857880425096\\
10.5323333333333	-0.0279637403778484\\
10.7075000000000	-0.0712335794389196\\
10.8826666666667	-0.0423578503180165\\
11.0578333333333	0.0260101161477041\\
11.2330000000000	0.0648127751195155\\
11.4081666666667	0.0380242664009473\\
11.5833333333333	-0.0242910926975223\\
11.7585000000000	-0.0592237766286314\\
11.9336666666667	-0.0342647218616170\\
12.1088333333333	0.0227681930675136\\
12.2840000000000	0.0543279474847582\\
12.4591666666667	0.0309837653497667\\
12.6343333333333	-0.0214106875452214\\
12.8095000000000	-0.0500144190469171\\
12.9846666666667	-0.0281048492252419\\
13.1598333333333	0.0201938246166740\\
13.3350000000000	0.0461937183198131\\
13.5101666666667	0.0255660714985579\\
13.6853333333333	-0.0190974213006973\\
13.8605000000000	-0.0427929456721573\\
14.0356666666667	-0.0233169035366870\\
14.2108333333333	0.0181049156628067\\
14.3860000000000	0.0397522590530716\\
14.5611666666667	0.0213158191766362\\
14.7363333333333	-0.0172025646154419\\
14.9115000000000	-0.0370221240117467\\
15.0866666666667	-0.0195284073681881\\
15.2618333333333	0.0163789125365243\\
15.4370000000000	0.0345612842248610\\
15.6121666666667	0.0179259949054990\\
15.7873333333333	-0.0156243108500673\\
15.7873333333333	-0.0323351227778032\\
15.9625000000000	-0.0164845107999103\\
16.1376666666667	0.0149306087005604\\
16.3128333333333	0.0303144450224117\\
16.4880000000000	0.0151836570472996\\
16.6631666666667	-0.0142908521370980\\
16.8383333333333	-0.0284744641149260\\
17.0135000000000	-0.0140061995984115\\
17.1886666666667	0.0136990985984793\\
17.3638333333333	0.0267940468686119\\
17.5390000000000	0.0129374536636761\\
17.7141666666667	-0.0131502210001253\\
17.8893333333333	-0.0252550643706893\\
18.0645000000000	-0.0119648263321427\\
18.2396666666667	0.0126397938540680\\
18.4148333333333	0.0238419114506661\\
18.5900000000000	0.0110774887103302\\
18.7651666666667	-0.0121639611212849\\
18.9403333333333	-0.0225410784611265\\
19.1155000000000	-0.0102660721345143\\
19.2906666666667	0.0117193650977836\\
19.4658333333333	0.0213408376189503\\
19.6410000000000	0.00952245465547186\\
19.8161666666667	-0.0113030544248579\\
19.9913333333333	-0.0202309530076781\\
20.1665000000000	-0.00883955387903801\\
};
\addplot [color=mycolor2, mark=,only marks, mark options={solid, fill=mycolor2, mycolor2}, mark size=1.2,forget plot]
  table[row sep=crcr]{%
2.82500000000000	1.03569422497496\\
3.00016666666667	0.562870346458428\\	
3.17533333333333	-0.309983612352802\\
3.35050000000000	-0.732086429343745\\
3.52566666666667	-0.415663830728522\\
3.70083333333333	0.219577064671528\\
3.87600000000000	0.545034418642228\\
4.05116666666667	0.318213636393385\\
4.22633333333333	-0.164641871131227\\
4.40150000000000	-0.421702683360320\\
4.57666666666667	-0.250604161014085\\
4.75183333333333	0.128818999284308\\
4.92700000000000	0.336107314053942\\
5.10216666666667	0.201893017155436\\
5.27733333333333	-0.104156285262082\\
5.45250000000000	-0.274273255504553\\
5.62766666666667	-0.165699797332892\\
5.80283333333333	0.0864303015013751\\
5.97800000000000	0.228142097340753\\
6.15316666666667	0.138110767865376\\
6.32833333333333	-0.0732386525245371\\
6.50350000000000	-0.192804419523631\\
6.67866666666667	-0.116623455571499\\
6.85383333333333	0.0631347602304598\\
7.02900000000000	0.165129841220588\\
7.20416666666667	0.0995784703049657\\
7.37933333333333	-0.0552073695647797\\
7.55450000000000	-0.143046866267611\\
7.72966666666667	-0.0858422933974002\\
7.90483333333333	0.0488589042116346\\
8.08000000000000	0.125139657009441\\
8.25516666666667	0.0746191967599317\\
8.43033333333333	-0.0436849892004562\\
8.60550000000000	-0.110414442777737\\
8.78066666666667	-0.0653382079962591\\
8.95583333333333	0.0394035313436268\\
9.13100000000000	0.0981565190383809\\
9.30616666666667	0.0575807650166567\\
9.48133333333333	-0.0358132672837859\\
9.65650000000000	-0.0878416085899265\\
9.83166666666667	-0.0510349236197294\\
10.0068333333333	0.0327670540786548\\
10.1820000000000	0.0790776800272373\\
10.3571666666667	0.0454641059708631\\
10.5323333333333	-0.0301555595042581\\
10.7075000000000	-0.0715671274289268\\
10.8826666666667	-0.0406866178911222\\
11.0578333333333	0.0278958548810623\\
11.2330000000000	0.0650804543013669\\
11.4081666666667	0.0365608424269015\\
11.5833333333333	-0.0259243136504335\\
11.7585000000000	-0.0594386035573880\\
11.9336666666667	-0.0329753338822269\\
12.1088333333333	0.0241912220938151\\
12.2840000000000	0.0545000049867306\\
12.4591666666667	0.0298412523699079\\
12.6343333333333	-0.0226574374985869\\
12.8095000000000	-0.0501517035112643\\
12.9846666666667	-0.0270872637782828\\
13.1598333333333	0.0212916199263853\\
13.3350000000000	0.0463025264208545\\
13.5101666666667	0.0246554218469450\\
13.6853333333333	-0.0200685630378361\\
13.8605000000000	-0.0428783625532104\\
14.0356666666667	-0.0224984061647402\\
14.2108333333333	0.0189676674779990\\
14.3860000000000	0.0398183418402678\\
14.5611666666667	0.0205771563678919\\
14.7363333333333	-0.0179720690236476\\
14.9115000000000	-0.0370722011658984\\
15.0866666666667	-0.0188593136443622\\
15.2618333333333	0.0170677432568883\\
15.4370000000000	0.0345980390787167\\
15.6121666666667	0.0173177962086642\\
15.7873333333333	-0.0162430356217891\\
15.7873333333333	-0.0323607875218802\\
15.9625000000000	-0.0159298902779520\\
16.1376666666667	0.0154881071787104\\
16.3128333333333	0.0303308337858163\\
16.4880000000000	0.0146763560328545\\
16.6631666666667	-0.0147946766410715\\
16.8383333333333	-0.0284831049690394\\
17.0135000000000	-0.0135408833160462\\
17.1886666666667	0.0141556604503750\\
17.3638333333333	0.0267961891860137\\
17.5390000000000	0.0125095093671223\\
17.7141666666667	-0.0135650307607096\\
17.8893333333333	-0.0252517725942734\\
18.0645000000000	-0.0115702863576932\\
18.2396666666667	0.0130175715375531\\
18.4148333333333	0.0238340594754101\\
18.5900000000000	0.0107128892945577\\
18.7651666666667	-0.0125088017942856\\
18.9403333333333	-0.0225294187834087\\
19.1155000000000	-0.00992841015483423\\
19.2906666666667	0.0120348039293078\\
19.4658333333333	0.0213259894545053\\
19.6410000000000	0.00920908570607418\\
19.8161666666667	-0.0115921844635031\\
19.9913333333333	-0.0202134549601586\\
20.1665000000000	-0.00854816914401513\\
};
\addplot [color=mycolor3, mark=, mark options={solid, fill=mycolor3, mycolor3}, forget plot]
  table[row sep=crcr]{%
2.82500000000000	1\\
3.00016666666667	0.886637593086455\\	
3.17533333333333	0.791513377600246\\
3.35050000000000	0.710914970560916\\
3.52566666666667	0.642028644838645\\
3.70083333333333	0.582690382614642\\
3.87600000000000	0.531213823044834\\
4.05116666666667	0.486269111113165\\
4.22633333333333	0.446796137532788\\
4.40150000000000	0.411941448657514\\
4.57666666666667	0.381011723866112\\
4.75183333333333	0.353439028578589\\
4.92700000000000	0.328754555705507\\
5.10216666666667	0.306568565764226\\
5.27733333333333	0.286554908046617\\
5.45250000000000	0.268438965085696\\
5.62766666666667	0.251988181732569\\
5.80283333333333	0.237004564428136\\
5.97800000000000	0.223318695817977\\
6.15316666666667	0.210784924673926\\
6.32833333333333	0.199277474579164\\
6.50350000000000	0.188687276154221\\
6.67866666666667	0.178919373058222\\
6.85383333333333	0.169890785993320\\
7.02900000000000	0.161528744572032\\
7.20416666666667	0.153769216385271\\
7.37933333333333	0.146555677520196\\
7.55450000000000	0.139838080272345\\
7.72966666666667	0.133571982717318\\
7.90483333333333	0.127717811773928\\
8.08000000000000	0.122240236864033\\
8.25516666666667	0.117107635598780\\
8.43033333333333	0.112291636356722\\
8.60550000000000	0.107766725362861\\
8.78066666666667	0.103509908079620\\
8.95583333333333	0.0995004164963571\\
9.13100000000000	0.0957194553423411\\
9.30616666666667	0.0921499814183409\\
9.48133333333333	0.0887765111980363\\
9.65650000000000	0.0855849526347308\\
9.83166666666667	0.0825624577546408\\
10.0068333333333	0.0796972931518176\\
10.1820000000000	0.0769787259425114\\
10.3571666666667	0.0743969231053111\\
10.5323333333333	0.0719428624411483\\
10.7075000000000	0.0696082536450843\\
10.8826666666667	0.0673854681984713\\
11.0578333333333	0.0652674769727060\\
11.2330000000000	0.0632477945901677\\
11.4081666666667	0.0613204297187932\\
11.5833333333333	0.0594798405879613\\
11.7585000000000	0.0577208951081393\\
11.9336666666667	0.0560388350577159\\
12.1088333333333	0.0544292438697891\\
12.2840000000000	0.0528880176112054\\
12.4591666666667	0.0514113387973674\\
12.6343333333333	0.0499956527304935\\
12.8095000000000	0.0486376460871829\\
12.9846666666667	0.0473342275141983\\
13.1598333333333	0.0460825100200626\\
13.3350000000000	0.0448797949750095\\
13.5101666666667	0.0437235575535586\\
13.6853333333333	0.0426114334729547\\
13.8605000000000	0.0415412068972923\\
14.0356666666667	0.0405107993916810\\
14.2108333333333	0.0395182598235506\\
14.3860000000000	0.0385617551194066\\
14.5611666666667	0.0376395617952098\\
14.7363333333333	0.0367500581872564\\
14.9115000000000	0.0358917173181184\\
15.0866666666667	0.0350631003390046\\
15.2618333333333	0.0342628504959213\\
15.4370000000000	0.0334896875723573\\
15.6121666666667	0.0327424027659637\\
15.7873333333333	0.0320198539609214\\
15.7873333333333	0.0313209613614525\\
15.9625000000000	0.0306447034552869\\
16.1376666666667	0.0299901132788946\\
16.3128333333333	0.0293562749589721\\
16.4880000000000	0.0287423205070766\\
16.6631666666667	0.0281474268464490\\
16.8383333333333	0.0275708130520016\\
17.0135000000000	0.0270117377861814\\
17.1886666666667	0.0264694969149821\\
17.3638333333333	0.0259434212897818\\
17.5390000000000	0.0254328746819571\\
17.7141666666667	0.0249372518583624\\
17.8893333333333	0.0244559767868013\\
18.0645000000000	0.0239885009615528\\
18.2396666666667	0.0235343018398577\\
18.4148333333333	0.0230928813810429\\
18.5900000000000	0.0226637646806541\\
18.7651666666667	0.0222464986925997\\
18.9403333333333	0.0218406510328857\\
19.1155000000000	0.0214458088590388\\
19.2906666666667	0.0210615778197975\\
19.4658333333333	0.0206875810700760\\
19.6410000000000	0.0203234583466096\\
19.8161666666667	0.0199688651000443\\
19.9913333333333	0.0196234716795658\\
20.1665000000000	0.0192869625664662\\
};
\end{axis}

\begin{axis}[%
width=1.227\fwidth,
height=1.227\fheight,
at={(-0.16\fwidth,-0.135\fheight)},
scale only axis,
xmin=0,
xmax=1,
ymin=0,
ymax=1,
axis line style={draw=none},
ticks=none,
axis x line*=bottom,
axis y line*=left,
ticklabel style={/pgf/number format/fixed}
]
\end{axis}
\end{tikzpicture}%